\documentclass[journal, twoside]{IEEEtran}
\PassOptionsToPackage{ruled, linesnumbered}{algorithm2e}
\usepackage[T1]{fontenc}
\usepackage[latin9]{inputenc}
\usepackage{color}
\usepackage{mathrsfs}
\usepackage{algorithm2e}
\usepackage{amsmath}
\usepackage{amsthm}
\usepackage{amssymb}
\usepackage{graphicx}
\usepackage[bookmarks=true,bookmarksnumbered=true,bookmarksopen=true,bookmarksopenlevel=1,
 breaklinks=false,pdfborder={0 0 0},pdfborderstyle={},backref=false,colorlinks=false]
 {hyperref}
\hypersetup{pdftitle={Your Title},
 pdfauthor={Your Name},
 pdfpagelayout=OneColumn, pdfnewwindow=true, pdfstartview=XYZ, plainpages=false}

\makeatletter
\let\oldforeign@language\foreign@language
\DeclareRobustCommand{\foreign@language}[1]{%
  \lowercase{\oldforeign@language{#1}}}
\theoremstyle{plain}
\newtheorem{thm}{\protect\theoremname}

\usepackage{amsmath, amsfonts, amssymb, amsthm, cite}
\usepackage[caption=false,font=footnotesize]{subfig}
\newcommand{\herm}{^{\mathsf{H}}}
\newcommand{\trans}{^{\mathsf{T}}}

\DeclareMathOperator{\minimize}{minimize}
\DeclareMathOperator{\diag}{\mathsf{diag}}
\DeclareMathOperator{\maximize}{maximize}
\DeclareMathOperator{\st}{subject~to}

\allowdisplaybreaks

\usepackage{lipsum}  
\usepackage{balance}

\newtheorem{remark}{Remark}

\makeatother

\providecommand{\theoremname}{Theorem}

\begin{document}
\title{Beamforming Design for Secure RIS-Enabled ISAC: Passive RIS vs. Active RIS}
\author{Vaibhav~Kumar,~\IEEEmembership{Member,~IEEE,} and~Marwa~Chafii,~\IEEEmembership{Senior~Member,~IEEE}\thanks{This work was supported by the Center for Cyber Security under the New York University Abu Dhabi Research Institute under Award G1104.}\thanks{Vaibhav~Kumar and Marwa~Chafii are with the Wireless Research Lab, Engineering Division, New York University Abu Dhabi, UAE. Marwa Chafii is also with NYU WIRELESS, NYU Tandon School of Engineering, New York, USA (e-mail: vaibhav.kumar@ieee.org; marwa.chafii@nyu.edu).}\thanks{Part of the content of this paper appeared in the Proc. of the IEEE Global Communications Conference (GLOBECOM\textquoteright 23), Kuala Lumpur, Malaysia 4-8 Dec. 2023~\cite{23-GC-SCA}.}}
\markboth{IEEE Transactions on Wireless Communications}{Kumar \MakeLowercase{\textit{et al.}}: Beamforming Design for Secure RIS-Enabled ISAC: Passive RIS vs. Active RIS}
\IEEEpubid{}
\maketitle
\begin{abstract}
The forthcoming sixth-generation (6G) communications standard is anticipated to provide integrated sensing and communication (ISAC) as a fundamental service. These ISAC systems present unique security challenges because of the exposure of information-bearing signals to sensing targets, enabling them to potentially eavesdrop on sensitive communication information with the assistance of sophisticated receivers. Recently, reconfigurable intelligent surfaces (RISs) have shown promising results in enhancing the physical layer security of various wireless communication systems, including ISAC. However, the performance of conventional passive RIS (pRIS)-enabled systems are often limited due to multiplicative fading, which can be alleviated using active RIS (aIRS). In this paper, we consider the problem of beampattern gain maximization in a secure pRIS/aRIS-enabled ISAC system, subject to signal-to-interference-plus-noise ratio constraints at communication receivers, and information leakage constraints at an eavesdropping target. For the challenging non-convex problem of joint beamforming design at the base station and the pRIS/aRIS, we propose a novel successive convex approximation (SCA)-based method. Unlike the conventional alternating optimization (AO)-based methods, in the proposed SCA-based approach, all of the optimization variables are updated simultaneously in each iteration. The proposed method shows significant performance superiority for pRIS-aided ISAC system compared to a benchmark scheme using penalty-based AO method. Moreover, our simulation results also confirm that aRIS-aided system has a notably higher beampattern gain at the target compared to that offered by the pRIS-aided system for the same power budget. We also present a detailed complexity analysis and proof of convergence for the proposed SCA-based method.
\end{abstract}


\IEEEpeerreviewmaketitle{}

\section{Introduction\protect\label{sec:Introduction}}

\IEEEPARstart{A}{s} the commercial deployment of the fifth-generation (5G)-enabled wireless services are underway across the globe, the third-generation partnership project (3GPP) has recently frozen release 17 (Rel-17) of the 5G standard. This latest release focuses on enhancing multicast and broadcast services, standalone non-public networks (SNPNs), non-terrestrial networks (NTNs), and reduced capability (RedCap) services, among others~\cite{Rel17}. The researchers are now looking ahead for the development of next-generation wireless standard to create a smart and connected wireless ecosystem, that requires a paradigm shift to support high-accuracy sensing capability along with high-quality wireless connectivity~\cite{23-ComMag-Naofal}. In their latest release (Rel-19), 3GPP has formed a technical specification group (TSG) to study the use-case of such sensing-aided wireless systems for intrusion detection in an outdoor/indoor area, rainfall/flood monitoring, autonomous maneuvering and navigation, monitoring the trajectory of an unmanned aerial vehicle (UAV), collision avoidance in an UAV network via network-assisted sensing, healthcare monitoring, extended-reality (XR) streaming, and advanced driver assistance system (ADAS) among various other applications. Based on the 3GPP-TSG studies, it is envisioned that sensing will be offered as a basic functionality in the sixth-generation (6G) wireless communications standard, resulting in a recent surge of research interest in integrated sensing and communication (ISAC). Through the integration of communication and sensing capabilities in a single network infrastructure, ISAC can exploit common hardware, as well as spectral and signal processing framework; it thereby can enjoy integration as well as coordination gain~\cite{22-ComMag-MultiFunctional6G,23-ComMag-ISAC-ML,23-COMST-12Chanllenges}\@.

At a higher level, ISAC can be classified into two categories: i) \emph{device-free} ISAC, and ii) \emph{device-based} ISAC. The former is based on device-free sensing in which the sensing targets do not have any transceiver capability, while in the latter case, sensing is facilitated via device-based sensing where the sensing targets are capable of transmitting and/or receiving radio signals~\cite{22-COMST-ISAC-Limits}. However, ISAC systems can also be classified based on the design requirements including radar-centric design~\cite{19-SpMag-DFRC}, communication-centric design~\cite{11-ProcIEEE-CommCentric}, and joint design~\cite{23-SpMag-70Yrs}. A comprehensive overview of the signal processing techniques for these three types of aforementioned ISAC systems was presented in~\cite{21-JSTSP-SigProc}. Some of the notable works on different aspects of ISAC were presented in~\cite{23-ComMag-MI,23-JSAC-FD-ISAC,23-TWC-ChannelModel,23-TWC-SaaS}. The advantages of an ISAC system over a frequency-division sensing and communications (FDSAC) system were studied in~\cite{23-ComMag-MI}. Using a mutual-information (MI)-based framework, it was shown therein that ISAC has a broader \emph{sensing-communication rate region} compared to that of FDSAC system in both uplink and downlink scenarios. The problems of joint design of transmit and receive beamformers for power consumption minimization and sum rate maximization in a full-duplex (FD) ISAC system were discussed in~\cite{23-JSAC-FD-ISAC}. A detailed channel modeling for ISAC combining forward and backward scattering, and including the non-stationarity and correlation between communication and sensing links was presented in~\cite{23-TWC-ChannelModel}. In~\cite{23-TWC-SaaS}, the authors presented a resource allocation scheme in a unified ISAC framework considering fairness and comprehensiveness criteria for a scalable trade-off between sensing and communication quality-of-service (QoS).

In recent times, passive reconfigurable intelligent surface (pRIS) has gained considerable attention as a groundbreaking hardware technology that can further enhance the performance of a wireless communication system~\cite{Minor_1,Minor-2}, including that of ISAC. These pRISs are composed of software-controlled metasurfaces capable of reengineering the wireless propagation media by controlling the phase of the electromagnetic waves reflected from these surfaces~\cite{23-ComMag-AdaptiveRIS}. The authors in~\cite{23-SPM-SPChepuri} considered the integration of these two emerging technologies, i.e., pRIS and ISAC, and shown that joint sensing and communication (S\&C) designs are mostly beneficial when the two channels were coupled together, and that using pRIS can be advantageous in the presence of such channel coupling. Similarly, it was shown in~\cite{23-WC-ISAC-RIS} that the use of RIS in an ISAC system can result in higher accuracy, wider coverage, and ultra-reliable communication and sensing performance. The problem of joint waveform and passive beamforming design for a multi-user multiple-input single-output (MU-MISO) ISAC system to maximize the signal-to-interference-and-noise-ratio (SINR) for radar and minimizing the multi-user interference was considered in~\cite{23-SPL-RIS-ISAC-MIMO}, where the authors proposed a solution to the challenging non-convex optimization problem using block coordinate descent (BCD) method. Similarly in~\cite{23-TWC-CRB}, the problem of joint beamforming design for pRIS-aided ISAC system was considered under signal-to-noise ratio (SNR) or Cram\'er-Rao bound (CRB) constraint, and a solution has been obtained using BCD. The authors in~\cite{23-TWC-PDD-MM} considered the problem of power allocation and beamforming design in an FD pRIS-ISAC system, and proposed a solution using penalty-dual-decomposition (PDD) method and majorization-minimization (MM). A low-complexity method using alternate direction method of multipliers (ADMM) was also proposed therein. Moreover, beamforming designs for simultaneously transmitting and reflecting (STAR)-RIS-aided ISAC system with non-orthogonal multiple access (NOMA), and pRIS-ISAC with rate-splitting multiple access (RSMA) were discussed in~\cite{23-TWC-STAR-RIS-ISAC-NOMA} and~\cite{24-WCL-RSMA-RIS-ISAC}, respectively. It is noteworthy that the problem of beamforming design in~\cite{23-SPL-RIS-ISAC-MIMO,23-TWC-CRB,23-TWC-PDD-MM,23-TWC-STAR-RIS-ISAC-NOMA,24-WCL-RSMA-RIS-ISAC} was solved using AO-based approach which results in a notably suboptimal solution due to the intricate coupling between the decision variables. 

Although ISAC provides a higher design flexibility and superior performance compared to the FDSAC counterpart, it also posses unique security challenges since the information-bearing communication signals are exposed to the potentially eavesdropping sensing targets. However, physical layer security (PLS) is an effective countermeasure for this problem. For example, the problem of beamforming design for physical layer security in in a satellite-terrestrial network was presented in~\cite{R2-1,R2-2}. Similarly, in the context of ISAC, the authors in~\cite{24-TIFS-Ahmad} devised an efficient beamforming design for transmit power minimization problem using the notion of physical layer security in an FD ISAC system. Furthermore, RIS has also been shown to enhance the PLS of wireless communication systems~\cite{21-WC-RIS},~\cite{R2-4}. In this direction, the authors in~\cite{23-TVT-Secure-RIS-ISAC} consider the secure beamforming design in an pRIS-aided MU-MISO ISAC system, where they aim to maximize the radar SINR, subject to SINR constraints at communication users, and information leakage constraints at the malicious target. In particular, the challenging non-convex optimization problem therein was solved via alternating optimization (AO) using semi-definite relaxation (SDR), fractional programming (FP), and MM; it was shown therein that pRIS-aided system has notable performance superiority compared to its non-pRIS counterpart. The problem of secure beamforming design in a similar MU-MISO system was presented in~\cite{23-COMML-secure-RIS-ISAC-Pd} to minimize the maximum SINR at an eavesdropper, while maintaining a predefined SINR threshold at communication receivers, and a detection probability threshold at the target. An iterative solution to the non-convex beamforming design problem was then obtained using AO technique based on FP and SDR. Moreover, optimized beamforming design to maximize the achievable sum secrecy rate in an pRIS-aided MU-MISO ISAC system was obtained in~\cite{23-WCL-secure-RIS-ISAC-sumRate} with the help of AO, successive convex approximation (SCA), and SDR. A penalty-assisted AO algorithm was proposed to obtain a semi-closed-form solution using Lagrange duality and an MM algorithm in~\cite{23-TWC-Benchmark} to obtain a secure beamforming design in an pRIS-aided MU-MISO ISAC system. Furthermore, a deep reinforcement learning (DRL)-based secure beamforming design in an pRIS-enabled MU-MISO ISAC system was developed in~\cite{23-TVT-RIS-ISAC-DRL}. At the same time, it is also interesting to keep in mind that a RIS can also pose severe security threats in wireless communication via destructive beamforming~\cite{R2-3}. One can easily notice that the solution to the beamforming design problem in secure pRIS-aided ISAC system in~\cite{24-TIFS-Ahmad,23-TVT-Secure-RIS-ISAC,23-COMML-secure-RIS-ISAC-Pd,23-WCL-secure-RIS-ISAC-sumRate,23-TWC-Benchmark,23-TVT-RIS-ISAC-DRL} were obtained using AO-based approaches where obtaining a stationary solution is not guaranteed theoretically. 

Although pRISs have shown promising benefits in ISAC systems, one of the major drawbacks in the pRIS-aided systems is that its performance is often restricted due to \emph{multiplicative fading} effect. This effect becomes more pronounced when there exist strong direct links between the transmitter and receiver(s). To circumvent this problem, \emph{active }RIS (aRIS) has been recently proposed, which is capable of amplifying the reflected signals by virtue of the integrated amplifiers in the RIS elements~\cite{23-TCOM-active_vs_passive}. An overview of aRIS-enabled ISAC system design focusing on important issues like frequency-selective fading, cascaded channel estimation, and optimal RIS placement was presented in~\cite{23-NetMag-aRIS-ISAC}. The authors in~\cite{23-WCL-aRIS-ISAC-THz}\textcolor{black}{{} }considered the problem of maximizing the target illumination power in an aRIS-aided terahertz (THz) ISAC system with delay alignment modulation, where they proposed an optimal beamforming design using AO, quadratically constrained quadratic program (QCQP), SDR, and MM. The problem of maximizing the SINR of the echo signal from the target in an aRIS-enabled MU-MISO ISAC system was considered in~\cite{23-TVT-aRIS-ISAC-Working}, where the beamforming design is obtained using AO, BCD method, Dinkelbach's transform, and MM. Similarly, in~\cite{23-TCOM-aRIS-ISAC-EarlyAccess}, the authors formulated the problem of maximizing the radar SINR, subject to communication SINRs in an aRIS-aided MU-MISO ISAC system; the optimal beamforming design therein was obtained using MM, AO, and SDR. A secure beamforming design for aRIS-enabled MU-MISO multicast ISAC system to maximize the achievable sum secrecy rate was proposed using AO, FP, and MM in~\cite{23-TVT-aRIS-ISAC-Secure}. Similar to the case of pRIS-aided systems, the solution to the beamforming optimization problem in aRIS-aided ISAC systems in~\cite{23-TCOM-active_vs_passive,23-WCL-aRIS-ISAC-THz,23-TVT-aRIS-ISAC-Working,23-TCOM-aRIS-ISAC-EarlyAccess,23-TVT-aRIS-ISAC-Secure} were also obtained using AO-based approaches. 

It is worth mentioning that compared to the pRIS-enabled ISAC systems, there is a dearth of literature on beamforming design for physical layer security in aRIS-enabled ISAC. Nevertheless, the beamforming design problem in pRIS/aRIS-aided ISAC systems are particularly challenging due to coupling between optimization variables, and therefore, AO is generally used to simplify the problem~\cite{23-SPL-RIS-ISAC-MIMO,23-TWC-CRB,23-TWC-PDD-MM,23-TWC-STAR-RIS-ISAC-NOMA,24-WCL-RSMA-RIS-ISAC},~\cite{23-TVT-Secure-RIS-ISAC,23-COMML-secure-RIS-ISAC-Pd,23-WCL-secure-RIS-ISAC-sumRate,23-TWC-Benchmark,23-TVT-RIS-ISAC-DRL},~\cite{23-TCOM-active_vs_passive,23-WCL-aRIS-ISAC-THz,23-TVT-aRIS-ISAC-Working,23-TCOM-aRIS-ISAC-EarlyAccess,23-TVT-aRIS-ISAC-Secure}. However, AO-based methods cannot produce a high-performance solution due to complicated coupling between optimization variables, and a stationary point is not guaranteed to be obtained theoretically~\cite{21-WC-OptMethods}. To address this challenge, we have used a method based on SCA in~\cite{23-GC-SCA} for the secure beamforming in a pRIS-enabled ISAC system. In this approach, we concurrently update all design variables in each iteration obtain a high-quality solution. Therefore, in this paper we propose an SCA-based optimization method for a high-performance beamforming design in a secure pRIS/aRIS-enabled MU-MISO ISAC system. Moreover, different from most of the existing literature on aRIS-aided system design~\cite{23-WCL-aRIS-ISAC-THz,23-TCOM-aRIS-ISAC-EarlyAccess,23-TVT-aRIS-ISAC-Secure,23-TCOM-active_vs_passive}, where separate power budgets are usually considered for the BS and aRIS, we consider a joint power budget for them. The reasoning behind this approach is as follows: we assume that both the BS and aRIS have unlimited power supplies, and both are connected to a central controller via backhaul links. The central controller determines an overall power budget based on the incentives paid by the users. Then, depending on the solution of the optimization problem, the controller informs the BS and aRIS about the portion of the power budget allocated to each. This method is more practical and adaptable than managing separate power budgets for the BS and aRIS. The main contributions of the paper are listed as follows:
\begin{itemize}
\item We consider the problem of beamforming design in a secure aRIS/pRIS-enabled MU-MISO ISAC system. In the system, the BS transmits a linear superposition of communication and sensing signals, which provides extra degrees-of-freedom (DoF) for beamforming design, and also help enhancing the security of the ISAC system. Our aim is to maximize the beampattern gain at the eavesdropping target, while ensuring a minimum SINR at communication users, and information leakage constraints at the target.
\item The formulated beampattern maximization problem is non-convex due to intricate coupling of the optimization variables, i.e., the transmit beamforming vector, and the aRIS/pRIS beamforming vector. For the case of aRIS-aided system, we also consider optimal power splitting between the BS and aRIS. Unlike the conventional AO-based approach, we use a series of convex approximations to reformulate the original non-convex optimization problem into an equivalent second-order cone program (SOCP), which can be solved using off-the-shelf solvers. The proposed method results in a high-performance solution since all of the design variables are updated simultaneously in each iteration. A proof of convergence and a complexity analysis are also presented for the proposed SCA-based method.
\item Extensive numerical results are then presented to analyze the performance of the proposed SCA-based algorithm. We also show the dependence of the system performance on different system design parameters. Our results confirm the superiority of the proposed algorithm over the AO-based benchmark algorithm in the case of pRIS-aided ISAC system. The results also confirm that the aRIS-aided system has significant performance superiority over the pRIS-aided system, however, this performance superiority is achieved at the cost of higher computational complexity/problem-solving time and memory requirement.
\end{itemize}

\paragraph*{Notation}

Uppercase and lowercase bold letters are used to represent matrices and vectors, respectively. The vector space containing all $M\times N$ complex-valued/real-valued matrices is indicated by $\mathbb{C}^{M\times N}/\mathbb{R}^{M\times N}$. The notation $\mathbf{X}\trans$, $\mathbf{X}\herm$, $\|\mathbf{X}\|/\|\mathbf{x}\|$, $\Re{\mathbf{X}}$, and $\Im{\mathbf{X}}$ refers to the transpose, conjugate transpose, Frobenius/Euclidean norm, real component, and imaginary component of a complex-valued matrix $\mathbf{X}$. The absolute value of a complex number $x$ is denoted by $|x|$, and $\diag(\mathbf{x})$ represents the diagonal matrix with elements from vector $\mathbf{x}$. The Bachmann\textendash Landau notation is denoted by $\mathcal{O}(\cdot)$.

\subsubsection*{Organization}

The remainder of this paper is organized as follows: Section~\ref{sec:System-Model-and-Problem-Formulation} introduces the pRIS/aRIS ISAC system considered in the paper and we formulate the problem of optimal secure beamforming design. In Section~\ref{sec:Proposed-Solution}, we propose a high-performance solution to the challenging non-convex optimization problem using a series of convex approximations. We also present a proof of convergence, and a detailed complexity analysis therein for the proposed SCA-based method. We present extensive simulation results to analyze the performance of the proposed algorithm for the ISAC system under consideration in Section~\ref{sec:Simulation-Results-and-Discussion}. Conclusions are then presented in Section~\ref{sec:Conclusion}.

\section{System Model and Problem Formulation\protect\label{sec:System-Model-and-Problem-Formulation}}

In this section, we first provide a description of the pRIS/aRIS-enabled ISAC system model, and then formulate the beampattern maximization problem.

\begin{figure}[t]
\begin{centering}
\includegraphics[width=0.9\columnwidth]{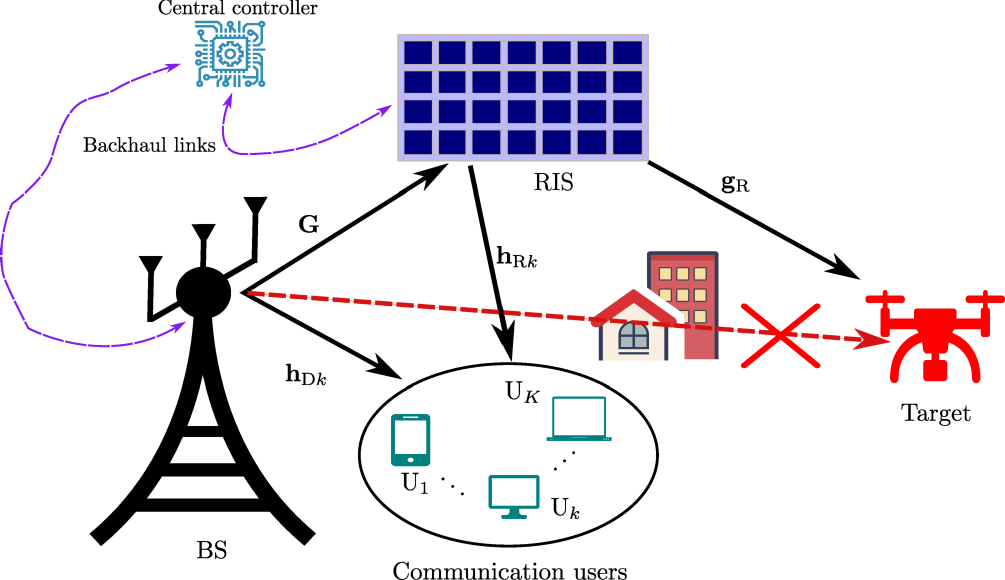}
\par\end{centering}
\caption{System model for secure RIS-enabled ISAC.}
\label{fig:SysMod}
\end{figure}

\subsection{System Model}

Consider the ISAC system shown in Fig.~\ref{fig:SysMod}, consisting of a dual-function radar-communication (DFRC) base station (BS) equipped with $L$ antennas, $K$ single-antenna communication users, one\footnote{Although our paper focuses on a single eavesdropping target scenario, extending the proposed algorithm to a system with multiple targets is straightforward. For multiple eavesdropping targets, the optimization objective can be formulated as maximizing the minimum beampattern gain among all targets. The SCA-based approach presented in the paper can be directly applied to this max-min optimization problem.} single-antenna eavesdropping target, and one passive/active RIS equipped with $N$ passive/active reflecting elements. We denote the communication users by $\mathrm{U}_{k}$ where $k\in\mathcal{K}\triangleq\{1,2,\ldots,K\}$. It is assumed that the BS transmits a linear superposition of information and sensing signals for the purpose of joint sensing and communication. Therefore, we denote the transmitted signal from the BS as 
\[
\mathbf{s}=\sum\nolimits_{k\in\mathcal{K}}\mathbf{x}_{\mathrm{c},k}w_{\mathrm{c},k}+\sum\nolimits_{m\in\mathcal{M}}\mathbf{x}_{\mathrm{t},m}w_{\mathrm{t},m},
\]
where $w_{\mathrm{c},k}\ (\forall k\in\mathcal{K})$ denotes the information-bearing communication signal for $\mathrm{U}_{k}\ (\forall k\in\mathcal{K})$, $w_{\mathrm{t},m}$ denotes the $m$-th sensing signal, $m\in\mathcal{M}\triangleq\{1,2,\ldots M\}$ with $M$ being the total number of sensing signals used, $\mathbf{x}_{\mathrm{c},k}\in\mathbb{C}^{L\times1}$ denotes the beamforming vector associated with $w_{\mathrm{c},k}$, and $\mathbf{x}_{\mathrm{t},m}\in\mathbb{C}^{L\times1}$ represents the beamforming vector corresponding to $w_{\mathrm{t},m}$. It is assumed that $\mathbb{E}\{w_{\mathrm{c},k}\}=\mathbb{E}\{w_{\mathrm{t},m}\}=0\ \forall k\in\mathcal{K},m\in\mathcal{M}$, $\mathbb{E}\{|w_{\mathrm{c},k}|^{2}\}=\mathbb{E}\{|w_{\mathrm{t},m}|^{2}\}=1\ \forall k\in\mathcal{K},m\in\mathcal{M}$ and $\mathbb{E}\{w_{\mathrm{c},k}w_{\mathrm{r},m}\herm\}=0\ \forall k\in\mathcal{K},m\in\mathcal{M}$, i.e., the communication and the sensing signals are uncorrelated. The BS-RIS, BS-$\mathrm{U}_{k}$, RIS-$\mathrm{U}_{k}$, and RIS-target wireless links are denoted by $\mathbf{G}\in\mathbb{C}^{N\times L}$, $\mathbf{h}_{\mathrm{D},k}\in\mathbb{C}^{1\times L}$, $\mathbf{h}_{\mathrm{R},k}\in\mathbb{C}^{1\times N}$, and $\mathbf{g}_{\mathrm{R}}\in\mathbb{C}^{1\times N}$, respectively. Similar to~\cite{23-TWC-Benchmark}, we assume that the direct links between the BS and target do not exist due to heave fading/shadowing or blockage.\footnote{Although in this work, we assume that the direct links between the BS and target are blocked, the algorithm proposed in this paper is also valid (without any changes) for the case when direct links between the two exist.} Therefore, the signal received at $\mathrm{U}_{k}$ is given by 
\begin{align}
y_{k} & =\big(\mathbf{h}_{\mathrm{D},k}+\mathbf{h}_{\mathrm{R},k}\boldsymbol{\Theta}\mathbf{G}\big)\mathbf{s}+\mathbf{h}_{\mathrm{R},k}\boldsymbol{\Theta}\mathbf{z}_{\mathrm{I}}+z_{k}\nonumber \\
 & =\mathbf{h}_{k}\mathbf{s}+\mathbf{h}_{\mathrm{R},k}\boldsymbol{\Theta}\mathbf{z}_{\mathrm{I}}+z_{k},\label{eq:rxSignal-Uk}
\end{align}
where $\mathbf{h}_{k}\triangleq\mathbf{h}_{\mathrm{D},k}+\mathbf{h}_{\mathrm{R},k}\boldsymbol{\Theta}\mathbf{G}$, $\boldsymbol{\Theta}\triangleq\diag(\boldsymbol{\theta})$, $\boldsymbol{\theta}$ is the RIS beamforming vector defined as $\boldsymbol{\theta}\triangleq[\theta_{1},\theta_{2},\ldots,\theta_{N}]\trans$, $\theta_{n}=\beta_{n}\exp(j\phi_{n})\ \forall n\in\mathscr{N}\triangleq\{1,2,\ldots,N\}$, $\phi_{n}\in[0,2\pi)$, $\mathbf{z}_{\mathrm{I}}\sim\mathcal{CN}(\boldsymbol{0},\sigma_{\mathrm{I}}^{2}\mathbf{I})\in\mathbb{C}^{N\times1}$ is the dynamic noise at the RIS, and $z_{k}\sim\mathcal{CN}(0,\sigma_{k}^{2})$ is the additive white Gaussian noise (AWGN) at $\mathrm{U}_{k}$. Note that for the case of passive RIS, $\beta_{n}=1$ and $\sigma_{\mathrm{I}}^{2}=0$, while for the case of active RIS, $\beta_{n}\leq\beta_{\max}$ and $\sigma_{\mathrm{I}}^{2}>0$. Therefore, the SINR to decode $w_{\mathrm{c},k}$ at $\mathrm{U}_{k}$ is given by 
\begin{align}
\gamma_{\mathrm{c},k} & =|\mathbf{h}_{k}\mathbf{x}_{\mathrm{c},k}|^{2}\big(\sigma_{k}^{2}+\sum\nolimits_{k'\in\mathcal{K}\setminus\{k\}}|\mathbf{h}_{k}\mathbf{x}_{\mathrm{c},k'}|^{2}\nonumber \\
 & \qquad+\sum\nolimits_{m\in\mathcal{M}}|\mathbf{h}_{k}\mathbf{x}_{\mathrm{t},m}|^{2}+\sigma_{\mathrm{I}}^{2}\|\mathbf{h}_{\mathrm{R},k}\boldsymbol{\Theta}\|^{2}\big)^{-1}.\label{eq:sinrUk}
\end{align}

Similarly, the signal received at the target is given by 
\begin{equation}
y_{\mathrm{t}}=\mathbf{g}_{\mathrm{R}}\boldsymbol{\Theta}\mathbf{G}\mathbf{s}+\mathbf{g}_{\mathrm{R}}\boldsymbol{\Theta}\mathbf{z}_{\mathrm{I}}+z_{\mathrm{t}}=\mathbf{g}_{\mathrm{t}}\mathbf{s}+\mathbf{g}_{\mathrm{R}}\boldsymbol{\Theta}\mathbf{z}_{\mathrm{I}}+z_{\mathrm{t}},\label{eq:rxSignal-target}
\end{equation}
where $\mathbf{g}_{\mathrm{t}}\triangleq\mathbf{g}_{\mathrm{R}}\boldsymbol{\Theta}\mathbf{G}$, and $z_{\mathrm{t}}\sim\mathcal{CN}(0,\sigma_{\mathrm{t}}^{2})$ is the AWGN at the target. Hence, the SINR at the target to decode $w_{\mathrm{c},k}$ is given by 
\begin{align}
\gamma_{\mathrm{t},k} & =|\mathbf{g}_{\mathrm{t}}\mathbf{x}_{\mathrm{c},k}|^{2}\big(\sigma_{\mathrm{t}}^{2}+\sum\nolimits_{k'\in\mathcal{K}\setminus\{k\}}|\mathbf{g}_{\mathrm{t}}\mathbf{x}_{\mathrm{c},k'}|^{2}\nonumber \\
 & \qquad+\sum\nolimits_{m\in\mathcal{M}}|\mathbf{g}_{\mathrm{t}}\mathbf{x}_{\mathrm{t},m}|^{2}+\sigma_{\mathrm{I}}^{2}\|\mathbf{g}_{\mathrm{R}}\boldsymbol{\Theta}\|^{2}\big)^{-1}.\label{eq:sinrTk}
\end{align}

Similar to~\cite{23-TWC-Benchmark,23-SPL-RIS-ISAC-MIMO,23-TWC-CRB,23-TWC-PDD-MM,23-TWC-STAR-RIS-ISAC-NOMA,24-WCL-RSMA-RIS-ISAC,24-TIFS-Ahmad,23-TVT-Secure-RIS-ISAC,23-COMML-secure-RIS-ISAC-Pd,23-WCL-secure-RIS-ISAC-sumRate,23-TVT-RIS-ISAC-DRL,23-WCL-aRIS-ISAC-THz,23-TVT-aRIS-ISAC-Working,23-TCOM-aRIS-ISAC-EarlyAccess,23-TVT-aRIS-ISAC-Secure}, we assume that all of the channels and the target location are perfectly known at the BS.\footnote{Although, considering imperfect knowledge of channel state information (CSI) and target location uncertainty are more practical, in this paper, we are interested to know the theoretical upper bound on the system performance. A similar assumption was considered in~\cite{23-SPL-RIS-ISAC-MIMO,23-TWC-CRB,23-TWC-PDD-MM,23-TWC-STAR-RIS-ISAC-NOMA,24-WCL-RSMA-RIS-ISAC,24-TIFS-Ahmad,23-TVT-Secure-RIS-ISAC,23-COMML-secure-RIS-ISAC-Pd,23-WCL-secure-RIS-ISAC-sumRate,23-TVT-RIS-ISAC-DRL,23-WCL-aRIS-ISAC-THz,23-TVT-aRIS-ISAC-Working,23-TVT-aRIS-ISAC-Secure,23-TCOM-aRIS-ISAC-EarlyAccess}. Moreover, several methods for obtaining the CSI in an RIS-aided systems were presented in~\cite{22-JSTSP-RIS-Signal-Processing,22-COMST-RIS-Channel-Estimation}. We leave the case of imperfect CSI and target location as a topic for future investigation.} Therefore, the beampattern gain at the target is defined as~(c.f.~\cite{23-TWC-Benchmark})
\begin{align}
 & \mathscr{G}(\mathbf{X},\boldsymbol{\theta})=\mathbb{E}\{|\mathbf{\mathbf{g}_{\mathrm{t}}s}+\mathbf{g}_{\mathrm{R}}\boldsymbol{\Theta}\mathbf{z}_{\mathrm{I}}|^{2}\}\nonumber \\
=\  & \mathbb{E}\{|\mathbf{\mathbf{g}_{\mathrm{t}}s}|^{2}+|\mathbf{g}_{\mathrm{R}}\boldsymbol{\Theta}\mathbf{z}_{\mathrm{I}}|^{2}+2\Re\{\mathbf{\mathbf{g}_{\mathrm{t}}s}(\mathbf{g}_{\mathrm{R}}\boldsymbol{\Theta}\mathbf{z}_{\mathrm{I}})\herm\}\}\nonumber \\
=\  & \sum_{k\in\mathcal{K}}|\mathbf{g}_{\mathrm{t}}\mathbf{x}_{\mathrm{c},k}|^{2}+\sum_{m\in\mathcal{M}}|\mathbf{g}_{\mathrm{t}}\mathbf{x}_{\mathrm{t},m}|^{2}+\sigma_{\mathrm{I}}^{2}\|\mathbf{g}_{\mathrm{R}}\boldsymbol{\Theta}\|^{2},\label{eq:BPG-Def}
\end{align}
where $\mathbf{X}\triangleq[\mathbf{x}_{\mathrm{c},1},\ldots,\mathbf{x}_{\mathrm{c},K},\mathbf{x}_{\mathrm{t},1},\ldots,\mathbf{x}_{\mathrm{t},M}]\in\mathbb{C}^{L\times(K+M)}$.

\subsection{Problem Formulation}

Similar to~\cite{23-TWC-Benchmark}, in this work we are interested in maximizing the beampattern gain at the target\footnote{This is equivalent to maximizing the average received power at the target or maximizing the target illumination power~\cite{23-WCL-aRIS-ISAC-THz}. This in turn can improve the SINR of the echo signal received from the target at the BS, or the probability of detection of the target. }, while also maintaining a certain predefined QoS at the communication users. Moreover, since the target is a potential eavesdropper, we impose a tolerance limit on the information leakage at the target. Therefore, for the case of \emph{active} RIS, the problem of jointly designing the BS and RIS beamforming vectors to maximize the beampattern gain at the target is formulated as follows: 
\begin{subequations}
\label{eq:mainProb-aIRS}
\begin{eqnarray}
(\mathscr{P}1)\qquad & \underset{\mathbf{X},\boldsymbol{\theta}}{\maximize} & \mathscr{G}(\mathbf{X},\boldsymbol{\theta}),\label{eq:mainObj}\\
 & \st & \gamma_{\mathrm{c},k}\geq\Gamma_{\mathrm{c},k}\ \forall k\in\mathcal{K},\label{eq:commSINRC}\\
 &  & \gamma_{\mathrm{t},k}\leq\Gamma_{\mathrm{t},k}\ \forall k\in\mathcal{K},\label{eq:targetSINRC}\\
 &  & \mathcal{P}(\mathbf{X},\boldsymbol{\theta})\leq P_{\max},\label{eq:TPC}\\
 &  & |\theta_{n}|\leq\beta_{\max}\ \forall n\in\mathscr{N},\label{eq:aIRSC}
\end{eqnarray}
\end{subequations}
where $\mathcal{P}(\mathbf{X},\boldsymbol{\theta})$ is the total power consumption at the BS and the active RIS, which is given by 
\begin{equation}
\mathcal{P}(\mathbf{X},\boldsymbol{\theta})=\|\mathbf{X}\|^{2}+\|\boldsymbol{\Theta}\mathbf{G}\mathbf{X}\|^{2}+\sigma_{\mathrm{I}}^{2}\|\boldsymbol{\Theta}\|^{2}.\label{eq:powerConsumption}
\end{equation}
In the RHS of the~\eqref{eq:powerConsumption}, the first terms $\|\mathbf{X}\|^{2}$ quantifies the total power of the signal transmitted from the BS, the second terms quantifies the total power of the (useful) signal reflected from the active RIS after amplification, and the third term quantifies the power of the dynamic noise generated during the signal amplification at the active RIS. Note that a similar modeling for power consumption at the aRIS had been used in~\cite{23-WCL-aRIS-ISAC-THz,23-TVT-aRIS-ISAC-Working}. It is worth mentioning that in the case of separate power budget constraints at the BS and aRIS (similar to the case in~\cite{23-WCL-aRIS-ISAC-THz,23-TCOM-aRIS-ISAC-EarlyAccess,23-TVT-aRIS-ISAC-Secure,23-TCOM-active_vs_passive}), the constraint at the BS will be a convex constraint, and the non-convex constraint at the aRIS can be handled following the same approach as used in~Theorem~\ref{thm-4}. Moreover,~\eqref{eq:commSINRC} ensures that the SINR at $\mathrm{U}_{k}$ is greater than or equal to a predefined threshold $\Gamma_{\mathrm{c},k}$,~\eqref{eq:targetSINRC} represents the tolerance limit on the leakage of $\mathrm{U}_{k}$'s information to the target\footnote{As highlighted in~\cite{HeterogeneousRequirements}, employing information leakage constraints provides greater flexibility in resource allocation compared to using secrecy rate when dealing with diverse security requirements. This approach is particularly beneficial for various applications such as video streaming, email services, IoT devices, and online banking, each of which may require different levels of security. The use of information leakage thresholds allows for more precise control over the secrecy performance against potential eavesdroppers, enabling a more effective balance between overall system performance and security measures. Alternatively, if one opts to evaluate secrecy performance using different metrics\textemdash such as secrecy rate, secrecy outage probability, or covert communication rate\textemdash the optimization strategy would need to be tailored to the specific system configuration and the convexity of the constraint function in question.}. Additionally,~\eqref{eq:TPC} ensures that the total power consumed at the BS and the aRIS is within the available budget $P_{\max}$, and~\eqref{eq:aIRSC} enforces that the amplification offered by each of the aRIS element is less than or equal to the maximum allowed amplification gain $\beta_{\max}$.\footnote{The constraint on the maximum amplification gain, i.e.,~\eqref{eq:aIRSC} is often overlooked in many papers related to aRIS. In such instances, the optimization algorithm yields an unconstrained optimal value of $|\theta_{n}|$; this value may not be achievable in practical applications due to hardware limitations.}

Analogously, for the case of \emph{passive }RIS, the problem of joint design of BS and RIS beamforming to maximize the beampattern gain at the target is given by 
\begin{subequations}
\label{eq:mainProb-pIRS}
\begin{eqnarray}
(\mathscr{P}2)\qquad & \underset{\mathbf{X},\boldsymbol{\theta}}{\maximize} & \mathscr{G}(\mathbf{X},\boldsymbol{\theta}),\label{eq:mainObj-pIRS}\\
 & \st & \eqref{eq:commSINRC},\eqref{eq:targetSINRC},\label{eq:SINRC-pIRS}\\
 &  & \|\mathbf{X}\|^{2}\leq P_{\max},\label{eq:TPC-pIRS}\\
 &  & |\theta_{n}|^{2}=1\ \forall n\in\mathscr{N}.\label{eq:UMC-pIRS}
\end{eqnarray}
\end{subequations}
Note that in~$(\mathscr{P}2)$, the SINR constraints at the users, and the information leakage constraints at the target are jointly represented using~\eqref{eq:SINRC-pIRS}, the transmit power constraint for the case of pRIS is represent using~\eqref{eq:TPC-pIRS}, and~\eqref{eq:UMC-pIRS} represents the unit-modulus constraint for each of the pRIS element.

It is noteworthy that both~$(\mathscr{P}1)$ and~$(\mathscr{P}2)$ are non-convex due to the non-convex objective function and the non-convex SINR and information leakage constraints. Moreover,~\eqref{eq:TPC} in~$(\mathscr{P}1)$, and~\eqref{eq:UMC-pIRS} in~$(\mathscr{P}2)$ are also non-convex. Additionally, the intricate coupling between the design variables $\mathbf{X}$ and $\boldsymbol{\theta}$ makes the problem more challenging to solve.

 Hua et al.~\cite{23-TWC-Benchmark} introduced a penalty-assisted dual-loop AO-based algorithm for solving~$(\mathscr{P}2)$. Specifically, in the inner loop, auxiliary variables were updated through solving a QCQP, the BS beamformers were updated using a bisection search, and the RIS reflection vector was updated via MM. The outer loop was solely utilized for updating the penalty parameter. Despite the reformulation of the optimization problem in~\cite{23-TWC-Benchmark}, where the design variables were decoupled in the constraints, achieving a high-quality solution is not guaranteed with AO. Furthermore, the per-iteration complexity of the penalty-based solution in~\cite{23-TWC-Benchmark} was $\mathcal{O}\big(N^{3}\big)$. However, due to the use of bisection search and the dual-loop structure, a large number of iterations is necessary for convergence. Consequently, this leads to extended problem-solving time, and prematurely terminating iterations before the penalty terms approach zero may yield infeasible solutions. Furthermore, it is not straightforward to apply the penalty-based AO method for the case of active RIS. To overcome these shortcomings, in the next section, we proposed an SCA-based method where all of the design variables are updated simultaneously in each iteration resulting in a high-performance solution.

\section{Proposed Solution\protect\label{sec:Proposed-Solution}}

In this section, we proposed an SCA-based method to obtain a stationary solution to~$(\mathscr{P}1)$ and~$(\mathscr{P}2)$. We first deal with the problem in~$(\mathscr{P}1)$, and then discuss on how the solution obtained for $(\mathscr{P}1)$ can be applied to $(\mathscr{P}2)$ with minimal modifications.

Before proceeding further, we recall the following (in)equalities:
\begin{subequations}
\label{eq:InEq}
\begin{align}
\|\mathbf{u}\|^{2}\geq & \ 2\Re\{\mathbf{v}\herm\mathbf{u}\}-\|\mathbf{v}\|^{2},\label{eq:InEq1}\\
\Re\{\mathbf{u}\herm\mathbf{v}\}= & \ \frac{1}{4}\big(\|\mathbf{u}+\mathbf{v}\|^{2}-\|\mathbf{u}-\mathbf{v}\|^{2}\big),\label{eq:InEq2}\\
\Im\{\mathbf{u}\herm\mathbf{v}\}= & \ \frac{1}{4}\big(\|\mathbf{u}-j\mathbf{v}\|^{2}-\|\mathbf{u}+j\mathbf{v}\|^{2}\big),\label{eq:InEq3}
\end{align}
\end{subequations}
which hold for two arbitrary complex-valued vectors $\mathbf{u}$ and $\mathbf{v}$. Using~\eqref{eq:InEq}, we apply a series of convex approximations on~$(\mathscr{P}1)$ in the following subsection.

\subsection{Solution to~$(\mathscr{P}1)$}

Due to the coupling between $\mathbf{X}$ and $\boldsymbol{\theta}$, the objective $\mathscr{G}(\mathbf{X},\boldsymbol{\theta})$ in~\eqref{eq:mainObj} is neither convex nor concave. Owing to the maximization nature of~$(\mathscr{P}1)$, we obtain a \emph{concave lower bound} on $\mathscr{G}(\mathbf{X},\boldsymbol{\theta})$ as follows: 
\begin{thm}
\label{thm-1}A concave lower bound on $\mathscr{G}(\mathbf{X},\boldsymbol{\theta})$ can be given by 
\begin{align}
\mathscr{G}(\mathbf{X},\boldsymbol{\theta})\geq\  & \sigma_{\mathrm{I}}^{2}\!\sum_{n\in\mathscr{N}}\!\hat{f}_{n}(\theta_{n};\theta_{n}^{(i)})+\sum_{k\in\mathcal{K}}f_{k}(\mathbf{x}_{\mathrm{c},k},\boldsymbol{\theta};\mathbf{x}_{\mathrm{c},k}^{(i)},\boldsymbol{\theta}^{(i)})\nonumber \\
\  & +\!\!\!\sum_{m\in\mathcal{M}}f_{m}(\mathbf{x}_{\mathrm{t},m},\boldsymbol{\theta};\mathbf{x}_{\mathrm{t},m}^{(i)},\boldsymbol{\theta}^{(i)})\triangleq\mathscr{F}(\mathbf{X},\boldsymbol{\theta}).\!\label{eq:thm-1}
\end{align}
\end{thm}
\begin{IEEEproof}
See Appendix~\ref{sec:proof-1}. 
\end{IEEEproof}
Next, we turn our attention to the non-convex constraints in~\eqref{eq:commSINRC}, which can be reformulated using the following theorem. 
\begin{thm}
\label{thm-2}The non-convex constraint in~~\eqref{eq:commSINRC} can be transformed into a set of convex constraints as follows: 
\begin{subequations}
\label{eq:thm-2}
\begin{align}
 & \frac{1}{\Gamma_{k}}\bar{f}_{k}\big(\mathbf{x}_{\mathrm{c},k},\boldsymbol{\theta};\mathbf{x}_{\mathrm{c},k}^{(i)},\boldsymbol{\theta}^{(i)}\big)\nonumber \\
 & \geq\sigma_{k}^{2}+\sum_{k'\in\mathcal{K}\setminus\{k\}}\big(\wp_{\mathrm{c},k,k'}^{2}+\bar{\wp}_{\mathrm{c},k,k'}^{2}\big)\nonumber \\
 & +\sum_{m\in\mathcal{M}}\!\!\big(\wp_{\mathrm{t},k,m}^{2}+\bar{\wp}_{\mathrm{t},k,m}^{2}\big)+\sigma_{\mathrm{I}}^{2}\|\mathbf{h}_{\mathrm{R},k}\boldsymbol{\Theta}\|^{2}\ \forall k\in\mathcal{K},\label{eq:thm-2-1}\\
 & \wp_{\mathrm{c},k,k'}\!\geq\!\mu_{\mathrm{c},1,k,k'}(\mathbf{x}_{\mathrm{c},k'},\boldsymbol{\theta};\mathbf{x}_{\mathrm{c},k'}^{(i)},\boldsymbol{\theta}^{(i)})\ \forall k\in\mathcal{K},k'\!\in\!\mathcal{K}\!\setminus\!\{k\},\!\label{eq:thm-2-2}\\
 & \wp_{\mathrm{c},k,k'}\!\geq\!\mu_{\mathrm{c},2,k,k'}(\mathbf{x}_{\mathrm{c},k'},\boldsymbol{\theta};\mathbf{x}_{\mathrm{c},k'}^{(i)},\boldsymbol{\theta}^{(i)})\ \forall k\in\mathcal{K},k'\!\in\!\mathcal{K}\!\setminus\!\{k\},\!\label{eq:thm-2-3}\\
 & \bar{\wp}_{\mathrm{c},k,k'}\!\geq\!\bar{\mu}_{\mathrm{c},1,k,k'}(\mathbf{x}_{\mathrm{c},k'},\boldsymbol{\theta};\mathbf{x}_{\mathrm{c},k'}^{(i)},\boldsymbol{\theta}^{(i)})\ \forall k\in\mathcal{K},k'\!\in\!\mathcal{K}\!\setminus\!\{k\},\!\label{eq:thm-2-4}\\
 & \bar{\wp}_{\mathrm{c},k,k'}\!\geq\!\bar{\mu}_{\mathrm{c},2,k,k'}(\mathbf{x}_{\mathrm{c},k'},\boldsymbol{\theta};\mathbf{x}_{\mathrm{c},k'}^{(i)},\boldsymbol{\theta}^{(i)})\ \forall k\in\mathcal{K},k'\!\in\!\mathcal{K}\!\setminus\!\{k\},\!\label{eq:thm-2-5}\\
 & \wp_{\mathrm{t},k,m}\!\geq\!\mu_{\mathrm{r},1,k,m}(\mathbf{x}_{\mathrm{t},m},\boldsymbol{\theta};\mathbf{x}_{\mathrm{t},m}^{(i)},\boldsymbol{\theta}^{(i)})\ \forall k\in\mathcal{K},m\!\in\!\mathcal{M},\!\label{eq:thm-2-6}\\
 & \wp_{\mathrm{t},k,m}\!\geq\!\mu_{\mathrm{r},2,k,m}(\mathbf{x}_{\mathrm{t},m},\boldsymbol{\theta};\mathbf{x}_{\mathrm{t},m}^{(i)},\boldsymbol{\theta}^{(i)})\ \forall k\in\mathcal{K},m\!\in\!\mathcal{M},\!\label{eq:thm-2-7}\\
 & \bar{\wp}_{\mathrm{t},k,m}\!\geq\!\bar{\mu}_{\mathrm{r},1,k,m}(\mathbf{x}_{\mathrm{t},m},\boldsymbol{\theta};\mathbf{x}_{\mathrm{t},m}^{(i)},\boldsymbol{\theta}^{(i)})\ \forall k\in\mathcal{K},m\!\in\!\mathcal{M},\!\label{eq:thm-2-8}\\
 & \bar{\wp}_{\mathrm{t},k,m}\!\geq\!\bar{\mu}_{\mathrm{r},2,k,m}(\mathbf{x}_{\mathrm{t},m},\boldsymbol{\theta};\mathbf{x}_{\mathrm{t},m}^{(i)},\boldsymbol{\theta}^{(i)})\ \forall k\in\mathcal{K},m\!\in\!\mathcal{M}.\!\label{eq:thm-2-9}
\end{align}
\end{subequations}
\end{thm}
\begin{IEEEproof}
See Appendix~\ref{sec:proof-2}.
\end{IEEEproof}
We now transform the non-convex constraint in~\eqref{eq:targetSINRC}, with the aid of the following theorem. 
\begin{thm}
\label{thm-3}The non-convex constraints in~\eqref{eq:targetSINRC} can be equivalently written using the following set of constraints:
\begin{subequations}
\label{eq:thm-3}
\begin{align}
 & \sigma_{\mathrm{t}}^{2}+\!\sigma_{\mathrm{I}}^{2}\!\sum_{n\in\mathscr{N}}\!\hat{f}_{n}(\theta_{n};\theta_{n}^{(i)})+\sum_{k'\in\mathcal{K}\setminus\{k\}}f_{k}(\mathbf{x}_{\mathrm{c},k'},\boldsymbol{\theta};\mathbf{x}_{\mathrm{c},k'}^{(i)},\boldsymbol{\theta}^{(i)})\nonumber \\
 & +\!\!\sum_{m\in\mathcal{M}}f_{m}(\mathbf{x}_{\mathrm{t},m},\boldsymbol{\theta};\mathbf{x}_{\mathrm{t},m}^{(i)},\boldsymbol{\theta}^{(i)})\!\geq\frac{1}{\Gamma_{\mathrm{t},k}}\big(\tau_{\mathrm{c},k}^{2}+\bar{\tau}_{\mathrm{c},k}^{2}\big)\ \forall k\in\mathcal{K},\label{eq:thm-3-1}\\
 & \tau_{\mathrm{\mathrm{c},k}}\geq\eta_{\mathrm{c},1,k}\big(\mathbf{x}_{\mathrm{c},k},\boldsymbol{\theta};\mathbf{x}_{\mathrm{c},k}^{(i)},\boldsymbol{\theta}^{(i)}\big)\ \forall k\in\mathcal{K},\label{eq:thm-3-2}\\
 & \tau_{\mathrm{\mathrm{c},k}}\geq\eta_{\mathrm{c},2,k}\big(\mathbf{x}_{\mathrm{c},k},\boldsymbol{\theta};\mathbf{x}_{\mathrm{c},k}^{(i)},\boldsymbol{\theta}^{(i)}\big)\ \forall k\in\mathcal{K},\label{eq:thm-3-3}\\
 & \bar{\tau}_{\mathrm{\mathrm{c},k}}\geq\bar{\eta}_{\mathrm{c},1,k}\big(\mathbf{x}_{\mathrm{c},k},\boldsymbol{\theta};\mathbf{x}_{\mathrm{c},k}^{(i)},\boldsymbol{\theta}^{(i)}\big)\ \forall k\in\mathcal{K},\label{eq:thm-3-4}\\
 & \bar{\tau}_{\mathrm{\mathrm{c},k}}\geq\bar{\eta}_{\mathrm{c},2,k}\big(\mathbf{x}_{\mathrm{c},k},\boldsymbol{\theta};\mathbf{x}_{\mathrm{c},k}^{(i)},\boldsymbol{\theta}^{(i)}\big)\ \forall k\in\mathcal{K},.\label{eq:thm-3-5}
\end{align}
\end{subequations}
\end{thm}
\begin{IEEEproof}
See Appendix~\ref{sec:proof-3}. 
\end{IEEEproof}
Next, we turn our attention to tackle the non-convex constraint in~\eqref{eq:TPC}. One can convexify~\eqref{eq:TPC} using the following theorem.
\begin{thm}
\label{thm-4}The non-convex constraint in~\eqref{eq:TPC} can be transformed to the following convex constraints:
\begin{subequations}
\label{eq:thm-4}
\begin{align}
P_{\mathrm{\max}}\geq & \ \sum_{k\in\mathcal{K}}\|\mathbf{x}_{\mathrm{c},k}\|^{2}+\sum_{m\in\mathcal{M}}\|\mathbf{x}_{\mathrm{t},m}\|^{2}\nonumber \\
 & +\sum_{k\in\mathcal{K}}\sum_{n\in\mathscr{N}}(\varkappa_{\mathrm{c},k,n}^{2}+\bar{\varkappa}_{\mathrm{c},k,n}^{2})\nonumber \\
 & +\sum_{m\in\mathcal{M}}\sum_{n\in\mathscr{N}}(\varkappa_{\mathrm{t},m,n}^{2}+\bar{\varkappa}_{\mathrm{t},m,n}^{2})+\sigma_{\mathrm{I}}^{2}\sum_{n\in\mathscr{N}}|\theta_{n}|^{2},\label{eq:thm-4-1}\\
\varkappa_{\mathrm{c},k,n}\geq & \ \varpi_{\mathrm{c},1,k,n}\big(\mathbf{x}_{\mathrm{c},k},\theta_{n};\mathbf{x}_{\mathrm{c},k}^{(i)},\theta_{n}^{(i)}\big)\ \forall k\in\mathcal{K},\forall n\in\mathscr{N},\label{eq:thm-4-2}\\
\varkappa_{\mathrm{c},k,n}\geq & \ \varpi_{\mathrm{c},2,k,n}\big(\mathbf{x}_{\mathrm{c},k},\theta_{n};\mathbf{x}_{\mathrm{c},k}^{(i)},\theta_{n}^{(i)}\big)\ \forall k\in\mathcal{K},\forall n\in\mathscr{N},\label{eq:thm-4-3}\\
\bar{\varkappa}_{\mathrm{c},k,n}\geq & \ \bar{\varpi}_{\mathrm{c},1,k,n}\big(\mathbf{x}_{\mathrm{c},k},\theta_{n};\mathbf{x}_{\mathrm{c},k}^{(i)},\theta_{n}^{(i)}\big)\ \forall k\in\mathcal{K},\forall n\in\mathscr{N},\label{eq:thm-4-4}\\
\bar{\varkappa}_{\mathrm{c},k,n}\geq & \ \bar{\varpi}_{\mathrm{c},2,k,n}\big(\mathbf{x}_{\mathrm{c},k},\theta_{n};\mathbf{x}_{\mathrm{c},k}^{(i)},\theta_{n}^{(i)}\big)\ \forall k\in\mathcal{K},\forall n\in\mathscr{N},\label{eq:thm-4-5}\\
\varkappa_{\mathrm{t},k,n}\geq & \ \varpi_{\mathrm{t},1,k,n}\big(\mathbf{x}_{\mathrm{t},m},\theta_{n};\mathbf{x}_{\mathrm{t},m}^{(i)},\theta_{n}^{(i)}\big)\ \forall m\in\mathcal{M},\forall n\in\mathscr{N},\label{eq:thm-4-6}\\
\varkappa_{\mathrm{t},k,n}\geq & \ \varpi_{\mathrm{t},2,k,n}\big(\mathbf{x}_{\mathrm{t},m},\theta_{n};\mathbf{x}_{\mathrm{t},m}^{(i)},\theta_{n}^{(i)}\big)\ \forall m\in\mathcal{M},\forall n\in\mathscr{N},\label{eq:thm-4-7}\\
\bar{\varkappa}_{\mathrm{t},k,n}\geq & \ \bar{\varpi}_{\mathrm{t},1,k,n}\big(\mathbf{x}_{\mathrm{t},m},\theta_{n};\mathbf{x}_{\mathrm{t},m}^{(i)},\theta_{n}^{(i)}\big)\ \forall m\in\mathcal{M},\forall n\in\mathscr{N},\label{eq:thm-4-8}\\
\bar{\varkappa}_{\mathrm{t},k,n}\geq & \ \bar{\varpi}_{\mathrm{t},2,k,n}\big(\mathbf{x}_{\mathrm{t},m},\theta_{n};\mathbf{x}_{\mathrm{t},m}^{(i)},\theta_{n}^{(i)}\big)\ \forall m\in\mathcal{M},\forall n\in\mathscr{N}.\label{eq:thm-4-9}
\end{align}
\end{subequations}
\end{thm}
\begin{IEEEproof}
See Appendix~\ref{sec:proof-4}.
\end{IEEEproof}
We are now left with the constraints in~\eqref{eq:aIRSC}, which is a convex constraint. Therefore, we may summarize the equivalent transformation of the problem in~\eqref{eq:mainProb-aIRS} as follows:
\begin{equation}
(\mathscr{P}3)\quad\underset{\mathbf{X},\boldsymbol{\theta},\mathcal{S}}{\maximize}\big\{\mathscr{F}(\mathbf{X},\boldsymbol{\theta})\mid\eqref{thm-1},\eqref{thm-2},\eqref{thm-3},\eqref{thm-4},\eqref{eq:aIRSC}\big\},\label{eq:P3}
\end{equation}
 where $\mathcal{S}\triangleq\{\boldsymbol{\wp}_{\mathrm{c}},\bar{\boldsymbol{\wp}}_{\mathrm{c}},\boldsymbol{\wp}_{\mathrm{t}},\bar{\boldsymbol{\wp}}_{\mathrm{t}},\boldsymbol{\tau}_{\mathrm{c}},\bar{\boldsymbol{\tau}}_{\mathrm{c}},\boldsymbol{\varkappa}_{\mathrm{c}},\bar{\boldsymbol{\varkappa}}_{\mathrm{c}},\boldsymbol{\varkappa}_{\mathrm{t}},\bar{\boldsymbol{\varkappa}}_{\mathrm{t}}\}$, $\boldsymbol{\wp}_{\mathrm{c}}\triangleq\{\wp_{\mathrm{c},k,k'}\mid k\in\mathcal{K},k'\in\mathcal{K}\setminus\{k\}\}$, $\bar{\boldsymbol{\wp}}_{\mathrm{c}}\triangleq\{\bar{\wp}_{\mathrm{c},k,k'}\mid k\in\mathcal{K},k'\in\mathcal{K}\setminus\{k\}\}$, $\boldsymbol{\wp}_{\mathrm{t}}\triangleq\{\wp_{\mathrm{t},k,m}\mid k\in\mathcal{K},m\in\mathcal{M}\}$, $\bar{\boldsymbol{\wp}}_{\mathrm{t}}\triangleq\{\bar{\wp}_{\mathrm{t},k,m}\mid k\in\mathcal{K},m\in\mathcal{M}\}$, $\boldsymbol{\tau}_{\mathrm{c}}\triangleq\{\tau_{\mathrm{c},k}\mid k\in\mathcal{K}\}$, $\bar{\boldsymbol{\tau}}_{\mathrm{c}}\triangleq\{\bar{\tau}_{\mathrm{c},k}\mid k\in\mathcal{K}\}$, $\boldsymbol{\varkappa}_{\mathrm{c}}\triangleq\{\varkappa_{\mathrm{c},k,n}\mid k\in\mathcal{K},n\in\mathscr{N}\}$, $\boldsymbol{\varkappa}_{\mathrm{c}}\triangleq\{\varkappa_{\mathrm{c},k,n}\mid k\in\mathcal{K},n\in\mathscr{N}\}$, $\bar{\boldsymbol{\varkappa}}_{\mathrm{c}}\triangleq\{\bar{\varkappa}_{\mathrm{c},k,n}\mid k\in\mathcal{K},n\in\mathscr{N}\}$, $\boldsymbol{\varkappa}_{\mathrm{t}}\triangleq\{\varkappa_{\mathrm{t},m,n}\mid m\in\mathcal{M},n\in\mathscr{N}\}$, and $\bar{\boldsymbol{\varkappa}}_{\mathrm{t}}\triangleq\{\bar{\varkappa}_{\mathrm{t},m,n}\mid m\in\mathcal{M},n\in\mathscr{N}\}$. Since all of the constraints in~$(\mathscr{P}3)$ can be represented by second-order cones, the problem in~$(\mathscr{P}3)$ is an SOCP problem. This can be efficiently solved using off-the-shelf solvers, e.g., CVX~\cite{CVX} / CVXPY~\cite{CVXPY} and MOSEK~\cite{mosek}. The proposed SCA-based SOCP algorithm to solve~$(\mathscr{P}3)$ is outlined in \textbf{Algorithm~\ref{algoP3}}.

\begin{algorithm}[t]
\caption{Proposed SCA-based Method to Solve~$(\mathscr{P}3)$.}

\label{algoP3}

\KwIn{$\mathbf{X}^{(0)}$, $\boldsymbol{\theta}^{(0)}$}

$i\leftarrow0$\;

\Repeat{convergence }{

Solve~$(\mathscr{P}3)$ and denote the solution as $\mathbf{X}^{\star}$, $\boldsymbol{\theta}^{\star}$\;

Update: $\mathbf{X}^{(i+1)}\leftarrow\mathbf{X}^{\star}$, $\boldsymbol{\theta}^{(i+1)}\leftarrow\boldsymbol{\theta}^{\star}$\;

$i\leftarrow i+1$\;

}

\KwOut{$\mathbf{X}^{\star}$, $\boldsymbol{\theta}^{\star}$}
\end{algorithm}

\setcounter{remark}{0}

\begin{remark}\label{remark-1}

To run \textbf{Algorithm~\ref{algoP3}}, we need to find \emph{feasible} initial points for~$(\mathscr{P}3)$, which is not straightforward to obtain. In the following, we describe a practical way to obtain such points. Consider the following optimization problem: 
\begin{subequations}
\label{eq:active_inital_points}
\begin{eqnarray}
(\mathscr{P}4) & \!\!\!\!\underset{\mathbf{X},\boldsymbol{\theta},\boldsymbol{\delta}_{\mathrm{c}}\geq\boldsymbol{0},\boldsymbol{\delta}_{\mathrm{t}}\geq\boldsymbol{0}}{\minimize} & \!\!\sum_{k\in\mathcal{K}}(\delta_{\mathrm{c},k}+\delta_{\mathrm{t},k}),\label{eq:feasible_objective}\\
 & \!\!\!\!\st & \!\!\delta_{\mathrm{c},k}+\gamma_{\mathrm{c},k}\geq\Gamma_{\mathrm{c},k}\ \forall k\in\mathcal{K},\label{eq:feasible_commSINR}\\
 &  & \!\!\gamma_{\mathrm{t},k}\leq\Gamma_{\mathrm{t},k}+\delta_{\mathrm{t},k}\ \forall k\in\mathcal{K},\label{eq:feasible_targetSINR}\\
 &  & \!\!\eqref{eq:TPC},\eqref{eq:aIRSC},
\end{eqnarray}
\end{subequations}
where $\boldsymbol{\delta}_{\mathrm{c}}\triangleq\{\delta_{\mathrm{c},k}\mid k\in\mathcal{K}\}$, and $\boldsymbol{\delta}_{\mathrm{t}}\triangleq\{\delta_{\mathrm{t},k}\mid k\in\mathcal{K}\}$. Note that~$(\mathscr{P}4)$ is feasible for sufficiently large $\boldsymbol{\delta}_{\mathrm{c}}$ and $\boldsymbol{\delta}_{\mathrm{t}}$, and can be solved following a similar set of transformation used to transform~$(\mathscr{P}1)$ to~$(\mathscr{P}3)$, and then applying an iterative algorithm similar to that outlined in~\textbf{Algorithm~\ref{algoP3}}. If at the convergence $\boldsymbol{\delta}_{\mathrm{c}}=\boldsymbol{\delta}_{\mathrm{t}}=\boldsymbol{0}$, the problem in~$(\mathscr{P}3)$ (equivalently~$(\mathscr{P}1)$) is considered to be feasible. We then choose the final values of $\mathbf{X}$ and $\boldsymbol{\theta}$ in~$(\mathscr{P}4)$ as $\mathbf{X}^{(0)}$ and $\boldsymbol{\theta}^{(0)}$ for~\textbf{Algorithm~\ref{algoP3}}. However, we simply declare that the considered problem is infeasible and will not run~\textbf{Algorithm~\ref{algoP3}} if the objective $\sum_{k\in\mathcal{K}}(\delta_{\mathrm{c},k}+\delta_{\mathrm{t},k})$ is not zero at the convergence.\end{remark}

\subsection{Solution to~$(\mathscr{P}2)$}

We now propose an SCA-based method to solve~$(\mathscr{P}2)$, similar to that proposed for~$(\mathscr{P}1)$. Following the arguments similar to those for the case of aRIS,~$(\mathscr{P}2)$ can be equivalently written as follows: 
\begin{subequations}
\label{eq:passive-1}
\begin{align}
\underset{\mathbf{X},\boldsymbol{\theta},\tilde{\mathcal{S}}}{\maximize}\  & \mathscr{F}(\mathbf{X},\boldsymbol{\theta}),\label{eq:passive-1-1}\\
\st\  & \eqref{eq:thm-1},\eqref{eq:thm-2},\eqref{eq:TPC-pIRS},\eqref{eq:UMC-pIRS},\nonumber 
\end{align}
\end{subequations}
where $\tilde{\mathcal{S}}\triangleq\{\boldsymbol{\wp}_{\mathrm{c}},\bar{\boldsymbol{\wp}}_{\mathrm{c}},\boldsymbol{\wp}_{\mathrm{t}},\bar{\boldsymbol{\wp}}_{\mathrm{t}},\boldsymbol{\tau}_{\mathrm{c}},\bar{\boldsymbol{\tau}}_{\mathrm{c}}\}$. Recall that in the case of pRIS, $\sigma_{\mathrm{I}}^{2}=0$. In~\eqref{eq:passive-1}, the only non-convex constraints are those in~\eqref{eq:UMC-pIRS}, which are particularly difficult to handle. Keeping in mind the increasing problem size, we tackle~\eqref{eq:UMC-pIRS} by first replacing the equality constraints with \emph{convex inequality constraints}, and then forcing the constraints to be \emph{binding} by adding a regularization term in the objective. This results in the following optimization problem: 
\begin{subequations}
\label{eq:passive-2}
\begin{align}
\underset{\mathbf{X},\boldsymbol{\theta},\tilde{\mathcal{S}}}{\maximize} & \ \mathscr{F}(\mathbf{X},\boldsymbol{\theta})+\zeta\|\boldsymbol{\theta}\|^{2}\label{eq:passive-2-1}\\
\st & \ \eqref{eq:thm-1},\eqref{eq:thm-2},\eqref{eq:TPC-pIRS},\nonumber \\
 & \ |\theta_{n}|\leq1\ \forall n\in\mathscr{N},\label{eq:passive-2-3}
\end{align}
\end{subequations}
where $\zeta>0$ is the regularization parameter. One can show that for sufficiently large $\zeta$,~\eqref{eq:passive-2-3} is binding at the optimality when convergence is achieved. Note that due to the convexity of the term $\zeta\|\boldsymbol{\theta}\|^{2}$, the term in~\eqref{eq:passive-2-1} is non-concave. Therefore, we use~\eqref{eq:InEq1} to obtain a corresponding concave lower bound as follows: 
\begin{subequations}
\begin{eqnarray*}
(\mathscr{P}5)\quad & \underset{\mathbf{X},\boldsymbol{\theta},\tilde{\mathcal{S}}}{\maximize} & \mathscr{F}(\mathbf{X},\boldsymbol{\theta})\!+\!\zeta[2\Re\{\boldsymbol{\theta}^{(i)}\ \!\!\herm\boldsymbol{\theta}\}\!-\!\|\boldsymbol{\theta}^{(i)}\|^{2}],\\
 & \st & \eqref{eq:thm-1},\eqref{eq:thm-2},\eqref{eq:TPC-pIRS},\eqref{eq:passive-2-3}.
\end{eqnarray*}
\end{subequations}
It can be shown that all of the constraints in~$(\mathscr{P}5)$ can be represented by quadratic cones, and therefore~$(\mathscr{P}5)$ is an SOCP problem. The proposed SAC-based method to solve~$(\mathscr{P}5)$ is outlined in~\textbf{Algorithm~\ref{algoP5}}.

\begin{algorithm}[t]
\caption{Proposed SCA-based Method to Solve~$(\mathscr{P}5)$.}

\label{algoP5}

\KwIn{$\mathbf{X}^{(0)}$, $\boldsymbol{\theta}^{(0)}$}

$i\leftarrow0$\;

\Repeat{convergence }{

Solve~$(\mathscr{P}5)$ and denote the solution as $\mathbf{X}^{\star}$, $\boldsymbol{\theta}^{\star}$\;

Update: $\mathbf{X}^{(i+1)}\leftarrow\mathbf{X}^{\star}$, $\boldsymbol{\theta}^{(i+1)}\leftarrow\boldsymbol{\theta}^{\star}$\;

$i\leftarrow i+1$\;

}

\KwOut{$\mathbf{X}^{\star}$, $\boldsymbol{\theta}^{\star}$}
\end{algorithm}

\begin{remark}

To solve~$(\mathscr{P}5)$, the initial points $\mathbf{X}^{(0)}$ and $\boldsymbol{\theta}^{(0)}$ can be obtained following a similar routine as described in~\textbf{Remark}\textbf{\emph{~\ref{remark-1}}}\emph{.}

\end{remark}

\paragraph*{Convergence analysis}

Let $(\mathbf{X}^{(i)},\boldsymbol{\theta}^{(i)})$ denotes the optimal solution to~$(\mathscr{P}3)$ in the $i$-th iteration. Hence, $(\mathbf{X}^{(i)},\boldsymbol{\theta}^{(i)})$ is also a feasible solution to~$(\mathscr{P}3)$ in the $(i+1)$-th itaration, and therefore, $\mathscr{F}(\mathbf{X}^{(i+1)},\boldsymbol{\theta}^{(i+1)})\geq\mathscr{F}(\mathbf{X}^{(i)},\boldsymbol{\theta}^{(i)})$. This means that~\textbf{Algorithm~\ref{algoP3}} generates a non-decreasing objective sequence. It is trivial to check that the objective function $\mathscr{F}(\mathbf{X},\boldsymbol{\theta})$ is continuous and bounded from above for a finite $P_{\max}$. Since the feasible set is compact, the objective sequence must converge to a finite limit, i.e., $\lim_{i\to\infty}\mathscr{F}(\mathbf{X}^{(i)},\boldsymbol{\theta}^{(i)})=\mathscr{F}^{\ast}$. Define $\mathscr{S}\triangleq\{\mathbf{X},\boldsymbol{\theta}|\mathscr{F}(\mathbf{X},\boldsymbol{\theta})\leq\mathscr{F}^{\ast}\}$, then due to the continuity and the compactness of the feasible set, and the continuity of $\mathscr{F}(\mathbf{X},\boldsymbol{\theta})$, the set $\mathscr{S}$ is compact. Hence there exists a sequence $(\mathbf{X}^{(i_{j})},\boldsymbol{\theta}^{(i_{j})})$ which converges to the limiting point $(\mathbf{X}^{\ast},\boldsymbol{\theta}^{\ast})$, and due to the continuity of $\mathscr{F}(\mathbf{X},\boldsymbol{\theta})$, one must have $\mathscr{F}^{\ast}=\mathscr{F}(\mathbf{X}^{\ast},\boldsymbol{\theta}^{\ast})$. It is rather standard to prove that $\mathscr{F}(\mathbf{X}^{\ast},\boldsymbol{\theta}^{\ast})$ is a stationary solution to~$(\mathscr{P}3)$, and therefore omitted. Interested readers may refer to\cite[Sec. 2.7]{1996-NLP}. The convergence analysis for~\textbf{Algorithm~\ref{algoP5} }can be carried out in a similar fashion, and hence omitted here for brevity.

\paragraph*{Complexity analysis}

Here we first compute the per-iteration complexity of the proposed SCA-based method for aRIS-aided system in~Algorithm~\ref{algoP3}. The total number of real-valued optimization variables in~$(\mathscr{P}3)$ is given by $\mathsf{N}_{\mathrm{var}}=2\{K^{2}+K(L+M+N)+LM+N\}+1$. At the same time, the total number of second-order conic constraints in~$(\mathscr{P}3)$ is given by $\mathsf{N}_{\mathrm{cons}}=4K^{2}+2K(1+2M+2N)+4MN+N+2.$ Next by computing the size of each conic constraint in~$(\mathscr{P}3)$, we define $\mathsf{N}_{\mathrm{size}}=[L(K\!+\!M)\!+\!1]^{2}\!+\!K[2(K\!+\!M\!-\!1)\!+\!N\!+\!L\!+1]^{2}+4K(K-1)(2L+1)^{2}+4KM(2L+1)^{2}+K(3+L)^{2}+4K(2L+1)^{2}+[2L(K+M)+2N(K+M)+2N]^{2}+4KN(2L+1)^{2}+4MN(2L+1)^{2}+4N.$  Therefore, following the arguments in~\cite[Sec. 6.6.2]{ModernOptLectures}, the per-iteration computational complexity for the proposed SCA-based method in~Algorithm~\ref{algoP3} is given by $\mathcal{C}_{\mathrm{aRIS-SCA}}=(\mathsf{N}_{\mathrm{cons}}+1)^{0.5}\mathsf{N}_{\mathrm{var}}(\mathsf{N}_{\mathrm{var}}^{2}+\mathsf{N}_{\mathrm{cons}}+\mathsf{N}_{\mathrm{size}}).$ However, in a practical system, $N\gg\max\{K,L,M\}$, and therefore, the approximate complexity can be given by $\mathcal{C}_{\mathrm{aRIS-SCA}}\approx O(N^{3.5}).$ Following a similar line of argument, one can show that for the pRIS-enabled system, the per-iteration computational complexity of the proposed SCA-based method in~Algorithm~\ref{algoP5} is also well-approximated by $\mathcal{C}_{\mathrm{pRIS-SCA}}\approx O(N^{3.5}).$ However, it can be shown that the \emph{exact} computational complexity of the proposed algorithm for the aRIS-enabled system is higher than that for the pRIS-aided system. In the following section, we will show that the former outperforms the latter; this performance superiority is obtained at the cost of a more expensive hardware, and higher computational complexity.

Additionally, as discussed in~\cite[Sec. III-C]{23-TWC-Benchmark}, the per-iteration complexity of the AO-based benchmark scheme for the pRIS-aided system can be approximated as $O(N^{3})$. Although, the per-iteration complexity of the proposed SCA-based method is slightly higher than that of the benchmark scheme, our experiments show that the proposed method requires significantly fewer iterations for convergence. We will show in the following section that for the case of pRIS-enabled system, this in turn results in a significantly smaller problem-solving time for the SCA-based method.

\section{Simulation Results and Discussion\protect\label{sec:Simulation-Results-and-Discussion}}

\begin{figure*}[t]
\begin{minipage}[t]{0.32\textwidth}%
\begin{center}
\includegraphics[width=0.99\textwidth]{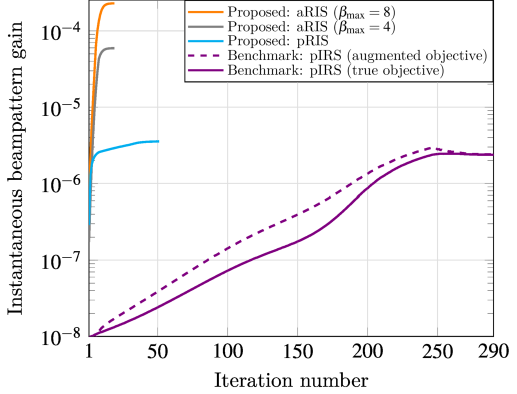}
\par\end{center}
\caption{Convergence results for $L=M=4$, $K=3$, $N=100$, $P_{\max}=40$~dBm, $\Gamma_{\mathrm{c}}=10$~dB, and $\Gamma_{\mathrm{t}}=0$~dB.}
\label{fig:convComp}%
\end{minipage}\hfill{}%
\begin{minipage}[t]{0.32\textwidth}%
\begin{center}
\includegraphics[width=0.99\textwidth]{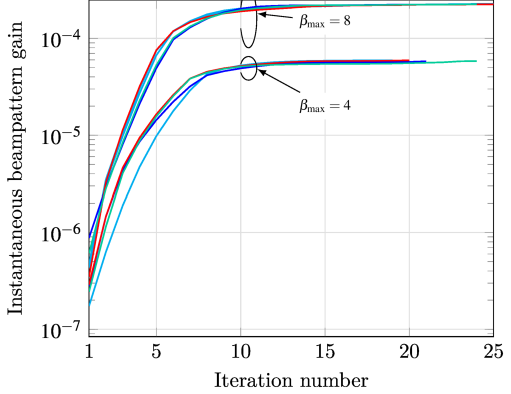}
\par\end{center}
\caption{Beampattern gain for $L=M=4$, $K=3$, $P_{\max}=40$~dBm, $N=100$, $\Gamma_{\mathrm{c}}=10$~dB, and $\Gamma_{\mathrm{t}}=0$~dB for different initial points.}
\label{fig_varyInitial}%
\end{minipage}\hfill{}%
\begin{minipage}[t]{0.32\textwidth}%
\begin{center}
\includegraphics[width=0.99\textwidth]{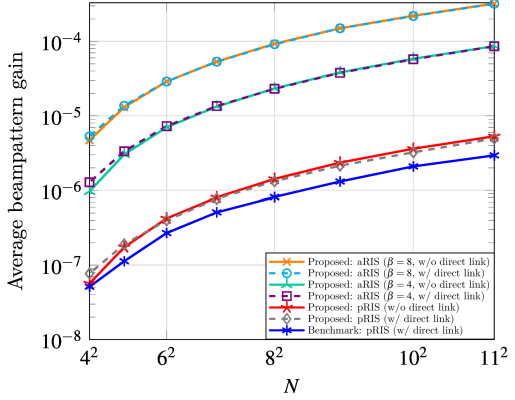}
\par\end{center}
\caption{Average beampattern gain for $L=4$, $K=3$, $P_{\max}=40$~dBm, $\Gamma_{\mathrm{c}}=10$~dB, $\Gamma_{\mathrm{t}}=0$~dB, and different values of $N$.}
\label{fig_varyN}%
\end{minipage}

\end{figure*}

In this section, we present comprehensive simulation results to investigate the performance of the proposed system and the corresponding algorithm.

\paragraph*{Simulation Setup}

\begin{figure*}[t]
\begin{minipage}[t]{0.32\textwidth}%
\begin{center}
\includegraphics[width=0.99\textwidth]{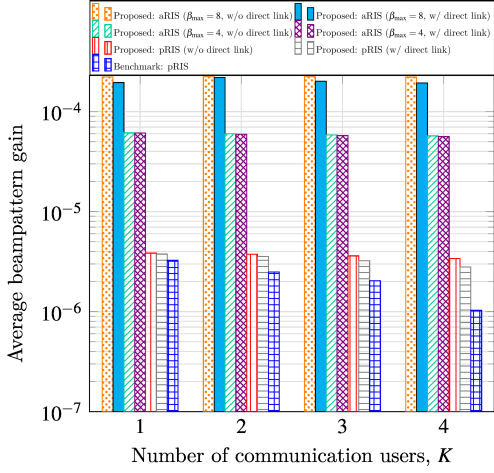}
\par\end{center}
\caption{Average beampattern gain for $L=M=4$, $P_{\max}=40$ dBm, $\Gamma_{\mathrm{c}}=10$~dB, $\Gamma_{\mathrm{t}}=0$~dB and different values of $K$.}
\label{fig:varyK}%
\end{minipage}\hfill{}%
\begin{minipage}[t]{0.32\textwidth}%
\begin{center}
\includegraphics[width=0.99\textwidth]{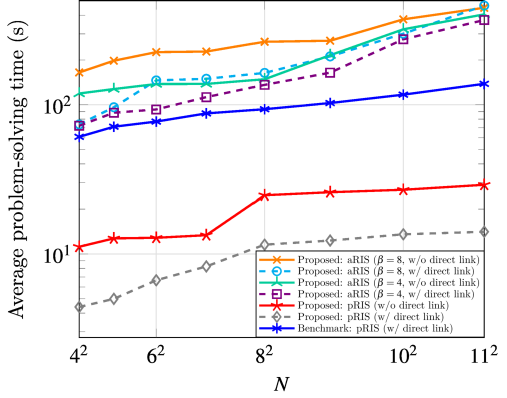}
\par\end{center}
\caption{Average problem-solving time for $L=4$, $K=3$, $P_{\max}=40$~dBm, $\Gamma_{\mathrm{c}}=10$~dB, $\Gamma_{\mathrm{t}}=0$~dB, and different values of $N$.}
\label{fig:Time}%
\end{minipage}\hfill{}%
\begin{minipage}[t]{0.32\textwidth}%
\begin{center}
\includegraphics[width=0.99\textwidth]{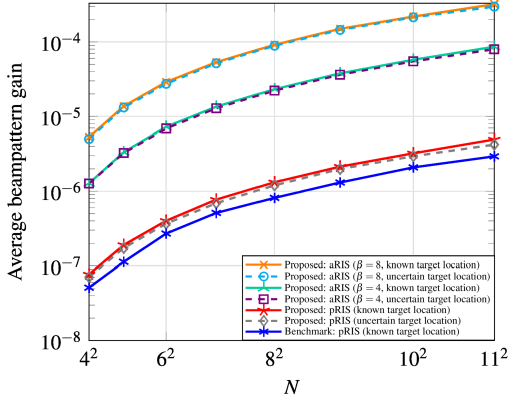}
\par\end{center}
\caption{Average beampattern gain for $L=4$, $K=3$, $P_{\max}=40$~dBm, $\Gamma_{\mathrm{c}}=10$~dB, $\Gamma_{\mathrm{t}}=0$~dB, and different values of $N$ for uncertain target location.}
\label{fig:UncertainTargetLocation}%
\end{minipage}
\end{figure*}

For the purpose of simulation, the location of different nodes (i.e., the BS, pRIS/aRIS, communication users, and the target) are assumed to be that same as that considered in~\cite{23-TWC-Benchmark}\textcolor{black}{.} Throughout this section, we assume $\Gamma_{\mathrm{c},k}=\Gamma_{\mathrm{c}}\ \forall k\in\mathcal{K}$, and $\Gamma_{\mathrm{t},k}=\Gamma_{\mathrm{t}}\ \forall k\in\mathcal{K}$. For the case of average beampattern gain, the averaging is performed over 100 independent channel realizations. Furthermore, in Figs.~\ref{fig:convComp} \textendash{} \ref{fig:UncertainTargetLocation}, the quantity in the y-axis has been represented in the logarithmic scale (with base 10) for the ease of readability. The simulations are performed on the NYUAD Jubail high performance computing cluster with AMD EPYC Rome CPU (128 cores \& 512GB RAM each) using Python v3.12.0 and MOSEK Fusion API for Python Rel.-10.1.21. 

\begin{remark}

\textcolor{black}{It is noteworthy that once a set of feasible points is obtained for $(\mathscr{P}3)$ (respectively$(\mathscr{P}5)$), it is guaranteed that the subsequent iterations of ~}\textbf{\textcolor{black}{Algorithm~\ref{algoP3} (}}\textcolor{black}{resp.~}\textbf{\textcolor{black}{Algorithm~\ref{algoP5}) }}\textcolor{black}{will return optimal/feasible solutions, because the optimal solution to $(\mathscr{P}3)$ (resp.~$(\mathscr{P}5)$) in the $i$-th iteration is also a feasible solution to $(\mathscr{P}3)$ (resp.~$(\mathscr{P}5)$) in the $(i+1)$-th iteration, respectively. However, sometime, due to the numerical issues, the optimization solver may return the ``solution status unknown'' error. To avoid this situation, we define the scaling factor $\varsigma$ as follows: 
\begin{align}
\varsigma & =\varepsilon/\max\{\max\{|\mathbf{G}|\},\max\{|\mathbf{h}_{\mathrm{D}1}|\},\ldots,\max\{|\mathbf{h}_{\mathrm{D}K}|\},\nonumber \\
 & \qquad\max\{|\mathbf{h}_{\mathrm{R}1}|\},\ldots,\max\{|\mathbf{h}_{\mathrm{R}K}|\},\max\{|\mathbf{g}_{\mathrm{R}}|\}\},\label{eq:scaling_factor}
\end{align}
where $\varepsilon=10$. Here $\max\{|\mathbf{X}|\}$ returns the element with largest absolute value in $\mathbf{X}$. Then we apply the following scaling:
\begin{align}
 & \{\mathbf{G},\mathbf{h}_{\mathrm{D}k},\mathbf{h}_{\mathrm{R}k},\mathbf{g}_{\mathrm{R}},\sigma_{k}^{2},\sigma_{\mathrm{t}}^{2},\sigma_{\mathrm{I}}^{2}\}\nonumber \\
 & \leftarrow\{\sqrt{\varsigma}\mathbf{G},\varsigma\mathbf{h}_{\mathrm{D}k},\sqrt{\varsigma}\mathbf{h}_{\mathrm{R}k},\sqrt{\varsigma}\mathbf{g}_{\mathrm{R}},\varsigma^{2}\sigma_{k}^{2},\varsigma^{2}\sigma_{\mathrm{t}}^{2},\varsigma^{2}\sigma_{\mathrm{I}}^{2}\}.\label{eq:channel_scaling}
\end{align}
Once~}\textbf{\textcolor{black}{Algorithm~\ref{algoP3} (}}\textcolor{black}{resp.~}\textbf{\textcolor{black}{Algorithm~\ref{algoP5}) }}\textcolor{black}{converges, we descale the objective function to obtain the true beampattern gain. }

\end{remark}

\paragraph*{Convergence results}

In Fig.~\ref{fig:convComp}, we show the convergence behavior of the proposed algorithm for aRIS/pRIS-enabled system, and compare it with that of the benchmark scheme proposed in~\cite{23-TWC-Benchmark}. For the dual-loop-based AO scheme~\cite{23-TWC-Benchmark}, we set the inner-loop and outer-loop tolerance as $10^{-2}$ and $10^{-4}$, respectively. At the same time, for the SCA-based proposed algorithm, we set the convergence tolerance equal to $10^{-3}$. Moreover, each iteration in Fig.~\ref{fig:convComp} for the case of AO-based benchmark scheme represents one iteration of the inner loop in~\cite[Algorithm 1]{23-TWC-Benchmark}. It can be noted from the figure that for the given set of channels, the benchmark scheme with pRIS-enabled system takes 290 iterations to converge. On the other hand, the proposed SCA-based scheme takes 50 iterations for pRIS-aided system and nearly 20 iterations for the aRIS-aided system. Additionally, a notable advantage of the proposed SCA-based method is that each iteration of the proposed method returns a feasible set of design variables, therefore the proposed algorithm can be stopped before convergence, if required. Contrary to this, the benchmark scheme returns a feasible set of optimization variables only in the final outer-loop iteration in~\cite[Algorithm 1]{23-TWC-Benchmark}. Therefore, in a rapidly changing environment with small coherence time, the benchmark method is not suitable where at least a suboptimal solution is required within a certain fraction of the channel's coherence time.

\paragraph*{Sensitivity to initialization}

In Fig.~\ref{fig_varyInitial}, we show the impact of initial values of $\mathbf{X}$ and $\boldsymbol{\theta}$. We use different feasible values of $(\mathbf{X},\boldsymbol{\theta})$ as initial points to solve $(\mathscr{P}4)$, and then use the corresponding converged values as initial points $(\mathbf{X}^{(0)},\boldsymbol{\theta}^{(0)})$ for the proposed SAC-based method in~\textbf{Algorithm~\ref{algoP3}}. It is evident from the figure that the proposed algorithm results in approximately the same beampattern gain at the convergence, irrespective of the initial points. In subsequent experiments, we conclude the iteration of the proposed method when either the relative change in the objective function is less than or equal to $10^{-3}$, or the iteration count reaches 50, whichever happens first. This ensures that the computational complexity remains within manageable limits.

\begin{figure*}[t]
\begin{minipage}[t]{0.32\textwidth}%
\begin{center}
\includegraphics[width=0.99\textwidth]{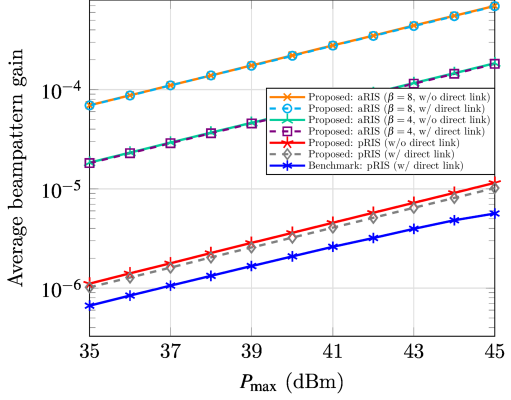}
\par\end{center}
\caption{Average beampattern gain for $L=4$, $K=3$, $N=100$, $\Gamma_{\mathrm{c}}=10$~dB, $\Gamma_{\mathrm{t}}=0$~dB, and different values of $P_{\max}$.}
\label{fig:varyPmax}%
\end{minipage}\hfill{}%
\begin{minipage}[t]{0.32\textwidth}%
\begin{center}
\includegraphics[width=0.99\textwidth]{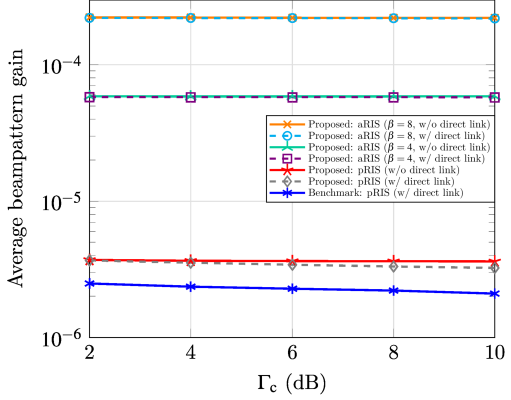}
\par\end{center}
\caption{Average beampattern gain for $L=4$, $K=3$, $N=100$, $\Gamma_{\mathrm{t}}=0$~dB, $P_{\max}=40$~dBm, and different values of $\Gamma_{\mathrm{c}}$.}
\label{fig:varyGamma}%
\end{minipage}\hfill{}%
\begin{minipage}[t]{0.32\textwidth}%
\begin{center}
\includegraphics[width=0.99\textwidth]{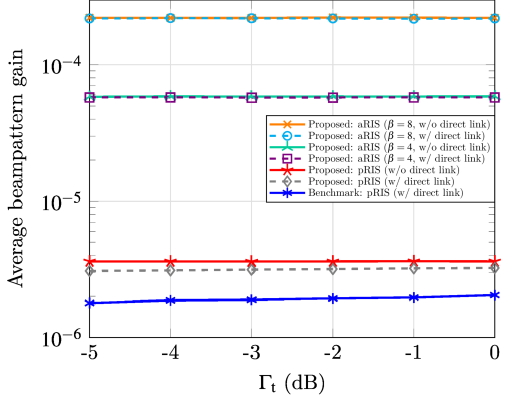}
\par\end{center}
\caption{Average beampattern gain for $L=4$, $K=3$, $N=100$, $\Gamma_{\mathrm{c}}=10$~dB, $P_{\max}=40$~dBm, and different values of $\Gamma_{\mathrm{t}}$.}
\label{fig:bvaryGammaR}%
\end{minipage}
\end{figure*}

\paragraph*{Impact of the number of RIS elements $N$}

Fig.~\ref{fig_varyN} shows the impact of the number of RIS elements on the average beampattern gain. An increase in the number of RIS elements increases the DoF at the RIS, allowing the RIS to perform highly focused beamforming. This in turn results in increasing beampattern gain with increasing $N$. Also, the superiority of the proposed method over the benchmark scheme can be easily noted from the figure, as the former results in a significantly higher average beampattern gain. Moreover, as $N$ increases, the impact of coupling between the design variables, i.e., $\mathbf{X}$ and $\boldsymbol{\theta}$ become more dominant, resulting in a highly suboptimal beampattern gain obtained via the AO-based benchmark method. On the other hand, the SCA-based method, where all of the design variables are updated simultaneously in each iteration, can efficiently handle the impact of coupling. In the figure, the legend ``w/o direct link'' and ``w/ direct link'' refer to the case of $\mathbf{h}_{\mathrm{D}k}=\boldsymbol{0}$ and $\mathbf{h}_{\mathrm{D}k}\neq0$, respectively. When the number of RIS elements is small, the system with direct link offers higher beampattern gain at the target compared to the w/o direct link counterpart because the communication QoS constraint are difficult to meet in the later case. However, when the number of reflecting elements increases, the system with direct link results in a smaller beampattern gain at the target. The reason for this non-trivial behavior is that when $N$ is large, the communication constraints can be met easily due to the large beamforming gain offered by the RIS, and since no direct link exists in the system, all of the power from the BS is focused toward the RIS, which in turn increases the total power received at the RIS (compared to the case with direct links, where some of the power from the BS is steered directly toward the communication users). Due to the larger incident power at RIS, the beampattern gain at the target increases. However, this effect is more pronounced in the case of pRIS-aided system.\textcolor{black}{{} } Additionally, the superiority of the aRIS-aided system over that of the pRIS-enabled system is also clearly evident from the figure, as the former effectively mitigates the impact of multiplicative fading.

\paragraph*{Impact of the power budget $P_{\max}$}

The impact of the power budget $P_{\max}$ on the average beampattern gain is shown in Fig.~\ref{fig:varyPmax}. For the case of pRIS-aided system, the inferiority of the AO-based scheme~\cite{23-TWC-Benchmark} is clearly evident from the figure. Moreover, the significant advantage of the aRIS-aided system over its pRIS-aided counterpart is also evident from the figure. The average beampattern gain for both the pRIS-aided and aRIS-assisted systems increases with an increase in $P_{\max}$, because of the reason that for a fixed $\Gamma_{\mathrm{c}}$, the required power from the BS to satisfy the communication SINR constraints remains fixed. So an increase in $P_{\max}$ results in increased amount of power steered toward the target, resulting in an increased beampattern gain at the target. 

\paragraph*{Impact of the communication SINR threshold $\Gamma_{\mathrm{c}}$}

Fig.~\ref{fig:varyGamma} shows the impact of communication SINR threshold, i.e., $\Gamma_{\mathrm{c}}$, on the average beampattern gain. As the communication SINR threshold increases, the SINR constraints become more stringent. This results in a reduced achievable average beampattern gain at the target, because a larger amount of power is required to satisfy the SINR constraint at the communication users. This in turn results in a reduced amount of surplus power for sensing the target. However, the reduction in the beampattern gain is more pronounced in the case of AO-based benchmark solution, highlighting its suboptimal system design. 

\paragraph*{Impact of the information leakage threshold $\Gamma_{\mathrm{t}}$}

The impact of varying information leakage threshold on the achievable average beampattern gain is shown in Fig.~\ref{fig:bvaryGammaR}. As the value of $\Gamma_{\mathrm{t}}$ increases, the information leakage constraints become less stringent, resulting in a higher average beampattern gain. This occurs because when the information leakage constraints are less stringent, more power can be directed toward the target, resulting in the increased beampattern gain. 

\paragraph*{Impact of the number of communication users}

The impact of the number of communication users, $K$, on the average beampattern gain of the aRIS/pRIS-enabled IRS system is shown in Fig.~\ref{fig:varyK}. One can note that increasing the number of communication users in the system results in a higher inter-user interference, and therefore, the meeting the communication SINR constraints becomes more difficult. This in turn needs a larger amount of power to be steered toward the communication users, which reduces the amount of surplus power to illuminate the target. This phenomenon leads to a reduction in the average beampattern gain at the target. However, we can easily note that this reduction in the beampattern gain with increasing number of communication users is more pronounced in the case of AO-based solution. 

\paragraph*{Average problem-solving time}

In Fig.~\ref{fig:Time}, we show the the average problem-solving time for different values of the number of RIS elements. As the value of $N$ increases, the size of the optimization problem to be solved also increases, resulting in an increasing average problem-solving time. However, for the case of pRIS-aided system, although the per-iteration complexity of the proposed SCA-based method is slightly higher than that of the AO-based benchmark scheme, the problem-solving time for the former is much smaller than that required by the latter. This happens because of the proposed method needs a significantly fewer number of iterations. On the other hand, for the aRIS-aided system, due to the increase in the size of the feasible set of the optimization variables, and a higher per-iteration complexity, the problem-solving time is the highest. For the exact same reason, the problem-solving time for the aRIS-aided system also increases with increasing $\beta_{\max}$. 

\paragraph*{Average beampattern gain for uncertain target location}

In Fig.~\ref{fig:UncertainTargetLocation}, we show the average beampattern gain for uncertain target location. To model the uncertain target location, we assume that $\varphi_{\mathrm{az}}\in[40^{\circ}-2.5^{\circ},40^{\circ}+2.5^{\circ}]$, and $\varphi_{\mathrm{el}}\in[-30^{\circ}-2.5^{\circ},-30^{\circ}+2.5^{\circ}]$; here $\varphi_{\mathrm{az}}$ and $\varphi_{\mathrm{el}}$ represent the target's azimuth and elevation angle, respectively. In this case, for the proposed SCA-based method, we assume that the BS first performs target location estimation with $5^{\circ}$ uncertainty in both azimuth and elevation angle, but ignores this estimation error for beamforming optimization, i.e., considers the estimated target location to be perfect. The robustness of the proposed algorithm against the target location uncertainty can be easily observed from the figure for both pRIS-aided and aRIS-aided systems due to the negligible performance degradation compared to the case of (perfectly) known target location. It is also noteworthy that for the case of pRIS-aided system, the proposed SCA-based method with uncertain target location significantly outperforms the benchmark system with known target location, which further highlights the drawbacks of AO-based algorithm in beamforming optimization in RIS-aided ISAC systems. 

\section{Conclusion\protect\label{sec:Conclusion}}

In this paper, we considered the problem of beamforming design in a pRIS/aRIS-enabled ISAC system with physical layer security. For the case of aRIS-enabled system, we considered a scenario where the beampattern gain at the eavesdropping target is maximized, subject to SINR constraint at the communication users, information leakage constraints at the target, optimal power allocation between the BS and aRIS, and maximum amplification constraint at the aRIS elements. On the other hand, for the case of pRIS-enabled system, the problem of beampattern gain maximization is constrained by SINR constraints, information leakage constraints, power budget constraint at the BS, and unit-modulus constraints for the pRIS elements. Different from the conventional AO-based approach, we proposed an SCA-based method where all of the design variables are updated simultaneously in each iteration. The proposed SCA-based method resulted in notable performance improvement in the case of pRIS-enabled system compared to that offered by a penalty-based AO method. With the aid of the proposed method, we also established the performance superiority of aRIS-aided system over the pRIS-assisted systems. This peformance superiority is achieved at the expense of higher hardware cost, and computational complexity/problem-solving time.

\appendices{}

\section{\protect\label{sec:proof-1}Proof of Theorem~\ref{thm-1}}

Using~\eqref{eq:BPG-Def}, we have 
\begin{align}
 & \mathscr{G}(\mathbf{X},\boldsymbol{\theta})\!=\!\sum_{k\in\mathcal{K}}|\mathbf{g}_{\mathrm{t}}\mathbf{x}_{\mathrm{c},k}|^{2}\!+\!\sum_{m\in\mathcal{M}}|\mathbf{g}_{\mathrm{t}}\mathbf{x}_{\mathrm{t},m}|^{2}\!+\!\sigma_{\mathrm{I}}^{2}\|\mathbf{g}_{\mathrm{R}}\boldsymbol{\Theta}\|^{2}\nonumber \\
=\  & \sum_{k\in\mathcal{K}}|\mathbf{g}_{\mathrm{t}}\mathbf{x}_{\mathrm{c},k}|^{2}+\sum_{m\in\mathcal{M}}|\mathbf{g}_{\mathrm{t}}\mathbf{x}_{\mathrm{t},m}|^{2}+\sigma_{\mathrm{I}}^{2}\sum_{n\in\mathscr{N}}|\theta_{n}g_{\mathrm{R}n}|^{2}\nonumber \\
\overset{(\texttt{a})}{\geq}\  & \sum_{k\in\mathcal{K}}\big[2\Re\{a_{\mathrm{c},k}^{(i)}\!\!\ \herm\mathbf{g}_{\mathrm{t}}\mathbf{x}_{\mathrm{c},k}\}-|a_{\mathrm{c},k}^{(i)}|^{2}\big]\nonumber \\
 & +\sum_{m\in\mathcal{M}}\big[2\Re\{a_{\mathrm{t},m}^{(i)}\!\!\ \herm\mathbf{g}_{\mathrm{t}}\mathbf{x}_{\mathrm{t},m}\}-|a_{\mathrm{t},m}^{(i)}|^{2}\big]\nonumber \\
 & +\sigma_{\mathrm{I}}^{2}\sum_{n\in\mathscr{N}}\big[2\Re\{\psi_{n}^{(i)}\!\!\ \herm g_{\mathrm{R}n}\theta_{n}\}-|\psi_{n}^{(i)}|^{2}\big]\nonumber \\
\overset{(\texttt{b})}{=}\  & \sum_{k\in\mathcal{K}}\!\Big[\frac{1}{2}\big\{\|a_{\mathrm{c},k}^{(i)}\mathbf{g}_{\mathrm{t}}\herm\!+\!\mathbf{x}_{\mathrm{c},k}\|^{2}\!-\!\|a_{\mathrm{c},k}^{(i)}\mathbf{g}_{\mathrm{t}}\herm\!-\!\mathbf{x}_{\mathrm{c},k}\|^{2}\big\}\!-\!|a_{\mathrm{c},k}^{(i)}|^{2}\Big]\nonumber \\
 & +\sum_{m\in\mathcal{M}}\Big[\frac{1}{2}\big\{\|a_{\mathrm{t},m}^{(i)}\mathbf{g}_{\mathrm{t}}\herm+\mathbf{x}_{\mathrm{t},m}\|^{2}-\!\|a_{\mathrm{t},m}^{(i)}\mathbf{g}_{\mathrm{t}}\herm\!-\mathbf{x}_{\mathrm{t},m}\|^{2}\big\}\nonumber \\
 & -\!|a_{\mathrm{t},m}^{(i)}|^{2}\Big]+\sigma_{\mathrm{I}}^{2}\sum_{n\in\mathscr{N}}\big[2\Re\{\psi_{n}^{(i)}\!\!\ \herm g_{\mathrm{R}n}\theta_{n}\}-|\psi_{n}^{(i)}|^{2}\big]\nonumber \\
\overset{(\texttt{c})}{\geq}\  & \sum_{k\in\mathcal{K}}\Big[\Re\big\{\mathbf{b}_{\mathrm{c},k}^{(i)}\!\!\ \herm\big[a_{\mathrm{c},k}^{(i)}\mathbf{g}_{\mathrm{t}}\herm+\mathbf{x}_{\mathrm{c},k}\big]\big\}-\frac{1}{2}\|\mathbf{b}_{\mathrm{c},k}^{(i)}\|^{2}\nonumber \\
 & \qquad\qquad-\frac{1}{2}\|a_{\mathrm{c},k}^{(i)}\mathbf{g}_{\mathrm{t}}\herm-\mathbf{x}_{\mathrm{c},k}\|^{2}-|a_{\mathrm{c},k}^{(i)}|^{2}\Big]\nonumber \\
 & +\sum_{m\in\mathcal{M}}\Big[\Re\big\{\mathbf{b}_{\mathrm{t},m}^{(i)}\!\!\ \herm\big[a_{\mathrm{t},m}^{(i)}\mathbf{g}_{\mathrm{t}}\herm+\mathbf{x}_{\mathrm{t},m}\big]\big\}-\frac{1}{2}\|\mathbf{b}_{\mathrm{t},m}^{(i)}\|^{2}\nonumber \\
 & \qquad\qquad-\frac{1}{2}\|a_{\mathrm{t},m}^{(i)}\mathbf{g}_{\mathrm{t}}\herm-\mathbf{x}_{\mathrm{t},k}\|^{2}-|a_{\mathrm{t},m}^{(i)}|^{2}\Big]\nonumber \\
 & +\sigma_{\mathrm{I}}^{2}\sum_{n\in\mathscr{N}}\big[2\Re\{\psi_{n}^{(i)}\!\!\ \herm g_{\mathrm{R}n}\theta_{n}\}-|\psi_{n}^{(i)}|^{2}\big]\nonumber \\
\triangleq & \sum_{k\in\mathcal{K}}f_{k}(\mathbf{x}_{\mathrm{c},k},\boldsymbol{\theta};\mathbf{x}_{\mathrm{c},k}^{(i)},\boldsymbol{\theta}^{(i)})\nonumber \\
 & +\!\!\sum_{m\in\mathcal{M}}f_{m}(\mathbf{x}_{\mathrm{t},m},\boldsymbol{\theta};\mathbf{x}_{\mathrm{t},m}^{(i)},\boldsymbol{\theta}^{(i)})\!+\!\sigma_{\mathrm{I}}^{2}\!\sum_{n\in\mathscr{N}}\!\hat{f}_{n}(\theta_{n};\theta_{n}^{(i)})\nonumber \\
\triangleq & \mathscr{F}(\mathbf{X},\boldsymbol{\theta}),\label{eq:objLowerBound}
\end{align}
where $\mathbf{x}_{\mathrm{c},k}^{(i)}$, $\mathbf{x}_{\mathrm{t},m}^{(i)}$ and $\boldsymbol{\theta}^{(i)}$ denote the value of $\mathbf{x}_{\mathrm{c},k}$, $\mathbf{x}_{\mathrm{t},m}$ and $\boldsymbol{\theta}$ in the $i$-th iteration of the SCA process, respectively, and $g_{\mathrm{R},n}$ is the $n$-th element of $\mathbf{g}_{\mathrm{R}}$. Moreover, $(\texttt{a})$ and $(\texttt{c})$ follow from~\eqref{eq:InEq1}, and $(\texttt{b})$ follows from~\eqref{eq:InEq2}. Additionally in~\eqref{eq:objLowerBound}, $a_{\mathrm{c},k}^{(i)}\triangleq\mathbf{g}_{\mathrm{t}}^{(i)}\mathbf{x}_{\mathrm{c},k}^{(i)}$, $a_{\mathrm{t},m}^{(i)}\triangleq\mathbf{g}_{\mathrm{t}}^{(i)}\mathbf{x}_{\mathrm{t},m}^{(i)}$, $\mathbf{b}_{\mathrm{c},k}^{(i)}\triangleq a_{\mathrm{c},k}^{(i)}\mathbf{g}_{\mathrm{t}}^{(i)}\!\!\ \herm+\mathbf{x}_{\mathrm{c},k}^{(i)}$, $\mathbf{b}_{\mathrm{t},m}^{(i)}\triangleq a_{\mathrm{t},m}^{(i)}\mathbf{g}_{\mathrm{t}}^{(i)}\!\!\ \herm+\mathbf{x}_{\mathrm{t},m}^{(i)}$, $\mathbf{g}_{\mathrm{t}}^{(i)}\triangleq\mathbf{g}_{\text{R}}\boldsymbol{\Theta}^{(i)}\mathbf{G}$, and $\psi_{n}^{(i)}\triangleq g_{\mathrm{R}n}\theta_{n}^{(i)}$. One can note that both $f_{k}(\mathbf{x}_{\mathrm{c},k},\boldsymbol{\theta};\mathbf{x}_{\mathrm{c},k}^{(i)},\boldsymbol{\theta}^{(i)})$ and $f_{m}(\mathbf{x}_{\mathrm{t},m},\boldsymbol{\theta};\mathbf{x}_{\mathrm{t},m}^{(i)},\boldsymbol{\theta}^{(i)})$ are jointly concave with respect to (w.r.t.) the BS beamforming vectors $\mathbf{x}_{\mathrm{c},k}$ and RIS beamforming vector $\boldsymbol{\theta}$,, and $\hat{f}_{n}(\theta_{n};\theta_{n}^{(i)})$ is concave w.r.t. $\boldsymbol{\theta}$. Hence,~\eqref{eq:objLowerBound} is jointly concave w.r.t. $\mathbf{X}$ and $\boldsymbol{\theta}$. This concludes the proof. 

\section{\protect\label{sec:proof-2}Proof of Theorem~\ref{thm-2}}

For any $k\in\mathcal{K}$, using~\eqref{eq:InEq}, we can equivalently represent~\eqref{eq:commSINRC} as 
\begin{subequations}
\label{eq:commSINR-1}
\begin{align}
\!\!\!\!\frac{1}{\Gamma_{k}}\big|\mathbf{h}_{k}\mathbf{x}_{k}\big|^{2} & \geq\sigma_{k}^{2}+\sum_{k'\in\mathcal{K}\setminus\{k\}}\big(\wp_{\mathrm{c},k,k'}^{2}+\bar{\wp}_{\mathrm{c},k,k'}^{2}\big)\nonumber \\
 & +\sum_{m\in\mathcal{M}}\big(\wp_{\mathrm{t},k,m}^{2}+\bar{\wp}_{\mathrm{t},k,m}^{2}\big)+\sigma_{\mathrm{I}}^{2}\|\mathbf{h}_{\mathrm{R},k}\boldsymbol{\Theta}\|^{2},\label{eq:cSINR-1-1}\\
\wp_{\mathrm{c},k,k'} & \geq\big|\Re\big\{\mathbf{h}_{k}\mathbf{x}_{\mathrm{c},k'}\big\}\big|\ \forall k'\in\mathcal{K}\setminus\{k\},\label{eq:cSINR-1-2}\\
\bar{\wp}_{\mathrm{c},k,k'} & \geq\big|\Im\big\{\mathbf{h}_{k}\mathbf{x}_{\mathrm{c},k'}\big\}\big|\ \forall k'\in\mathcal{K}\setminus\{k\},\label{eq:cSINR-1-3}\\
\wp_{\mathrm{t},k,m} & \geq\big|\Re\big\{\mathbf{h}_{k}\mathbf{x}_{\mathrm{t},m}\big\}\big|\ \forall m\in\mathcal{M},\label{eq:cSINR-1-4}\\
\bar{\wp}_{\mathrm{t},k,m} & \geq\big|\Im\big\{\mathbf{h}_{k}\mathbf{x}_{\mathrm{t},m}\big\}\big|\ \forall m\in\mathcal{M}.\label{eq:cSINR-1-5}
\end{align}
\end{subequations}
In~\eqref{eq:cSINR-1-1}, we only need to obtain a concave lower bound on the left-hand side (LHS), since the right-hand side (RHS) is already convex. Analogous to~\eqref{eq:objLowerBound}, this can be achieved as follows:
\begin{align}
\frac{1}{\Gamma_{k}}\big|\mathbf{h}_{k}\mathbf{x}_{\mathrm{c},k}\big|^{2} & \geq\frac{1}{\Gamma_{k}}\big[\Re\big\{\big(\mathbf{d}_{\mathrm{c},k}^{(i)}\!\!\ \herm\big)\big[c_{\mathrm{c},k}^{(i)}\mathbf{h}_{k}\herm+\mathbf{x}_{\mathrm{c},k}\big]\big\}\nonumber \\
 & \quad-\frac{1}{2}\|\mathbf{d}_{\mathrm{c},k}^{(i)}\|^{2}-\frac{1}{2}\|c_{\mathrm{c},k}^{(i)}\mathbf{h}_{k}\herm-\mathbf{x}_{\mathrm{c},k}\|^{2}-\big|c_{\mathrm{c},k}^{(i)}\big|^{2}\big]\nonumber \\
 & \triangleq\frac{1}{\Gamma_{k}}\bar{f}_{k}\big(\mathbf{x}_{\mathrm{c},k},\boldsymbol{\theta};\mathbf{x}_{\mathrm{c},k}^{(i)},\boldsymbol{\theta}^{(i)}\big),\label{eq:fBarDef}
\end{align}
where $c_{\mathrm{c},k}^{(i)}\triangleq\mathbf{h}_{k}^{(i)}\mathbf{x}_{\mathrm{c},k}^{(i)}$ and $\mathbf{d}_{\mathrm{c},k}^{(i)}\triangleq c_{\mathrm{c},k}^{(i)}\mathbf{h}_{k}^{(i)}\!\!\ \herm+\mathbf{x}_{\mathrm{c},k}^{(i)}$. Therefore, using~\eqref{eq:cSINR-1-1} and~\eqref{eq:fBarDef}, the constraint in~\eqref{eq:cSINR-1-1} can be transformed as~\eqref{eq:thm-2-1}.

Next, using the fact that $u\geq|v|$ iff $u\geq v$ or $u\geq|-v|$, and following~\eqref{eq:InEq2}, $\wp_{k\ell}$ in~\eqref{eq:cSINR-1-2} can be equivalently written as 
\begin{subequations}
\label{eq:tkl-1}
\begin{align}
\wp_{\mathrm{c},k,k'} & \geq\Re\big\{\mathbf{h}_{k}\mathbf{x}_{\mathrm{c},k'}\big\}\nonumber \\
 & =\frac{1}{4}\big(\|\mathbf{h}_{k}\herm+\mathbf{x}_{\mathrm{c},k'}\|^{2}-\|\mathbf{h}_{k}\herm-\mathbf{x}_{\mathrm{c},k'}\|^{2}\big),\label{eq:tkl-1-1}\\
\wp_{\mathrm{c},k,k'} & \geq-\Re\big\{\mathbf{h}_{k}\mathbf{x}_{\mathrm{c},k'}\big\}\nonumber \\
 & =\frac{1}{4}\big(\|\mathbf{h}_{k}\herm-\mathbf{x}_{\mathrm{c},k'}\|^{2}-\|\mathbf{h}_{k}\herm+\mathbf{x}_{\mathrm{c},k'}\|^{2}\big).\label{eq:tkl-1-2}
\end{align}
\end{subequations}
Note that the RHSs of~\eqref{eq:tkl-1-1} and~\eqref{eq:tkl-1-2} are non-convex due to the negative quadratic terms $-\|\mathbf{h}_{k}\herm-\mathbf{x}_{\mathrm{c},k'}\|^{2}$ and $-\|\mathbf{h}_{k}\herm+\mathbf{x}_{\mathrm{c},k'}\|^{2}$, respectively, while we need the RHSs to be convex. Therefore, we invoke the inequality in~\eqref{eq:InEq1} in~\eqref{eq:tkl-1-1} as follows:
\begin{align}
\wp_{\mathrm{c},k,k'} & \geq\frac{1}{4}\big[\|\mathbf{h}_{k}\herm+\mathbf{x}_{\mathrm{c},k'}\|^{2}-2\Re\big\{\big(\mathbf{h}_{k}^{(i)}-\mathbf{x}_{\mathrm{c},k'}^{(i)}\!\!\ \herm\big)\nonumber \\
 & \quad\qquad\times\big(\mathbf{h}_{k}\herm-\mathbf{x}_{\mathrm{c},k'}\big)\big\}+\|\mathbf{h}_{k}^{(i)}\!\!\ \herm-\mathbf{x}_{\mathrm{c},k'}^{(i)}\|^{2}\big]\nonumber \\
 & \triangleq\mu_{\mathrm{c},1,k,k'}\big(\mathbf{x}_{\mathrm{c},k'},\boldsymbol{\theta};\mathbf{x}_{\mathrm{c},k'}^{(i)},\boldsymbol{\theta}^{(i)}\big).\label{eq:mu_c1_k_kPrime}
\end{align}
One can note that $\mu_{\mathrm{c},1,k,k'}\big(\mathbf{x}_{\mathrm{c},k'},\boldsymbol{\theta};\mathbf{x}_{\mathrm{c},k'}^{(i)},\boldsymbol{\theta}^{(i)}\big)$ is convex due to the RHS of~\eqref{eq:mu_c1_k_kPrime} being jointly convex w.r.t. $\mathbf{x}_{\mathrm{c},k,k'}$ and $\boldsymbol{\theta}$. Following similar arguments, we can transform the inequality in~\eqref{eq:tkl-1-2} as 
\begin{align}
\wp_{\mathrm{c},k,k'} & \geq\frac{1}{4}\big[\|\mathbf{h}_{k}\herm-\mathbf{x}_{\mathrm{c},k'}\|^{2}-2\Re\big\{\big(\mathbf{h}_{k}^{(i)}+\mathbf{x}_{\mathrm{c},k'}^{(i)}\!\!\ \herm\big)\nonumber \\
 & \quad\qquad\times\big(\mathbf{h}_{k}\herm+\mathbf{x}_{\mathrm{c},k'}\big)\big\}+\|\mathbf{h}_{k}^{(i)}\!\!\ \herm+\mathbf{x}_{\mathrm{c},k'}^{(i)}\|^{2}\big]\nonumber \\
 & \triangleq\mu_{\mathrm{c},2,k,k'}\big(\mathbf{x}_{\mathrm{c},k'},\boldsymbol{\theta};\mathbf{x}_{\mathrm{c},k'}^{(i)},\boldsymbol{\theta}^{(n)}\big).\label{eq:mu_c2_k_kPrime}
\end{align}
Analogously,~\eqref{eq:cSINR-1-3} \textendash{} \eqref{eq:cSINR-1-5} can be transformed as 
\begin{align}
\bar{\wp}_{\mathrm{c},k,k'} & \geq\frac{1}{4}\big[\|\mathbf{h}_{k}\herm-j\mathbf{x}_{\mathrm{c},k'}\|^{2}-2\Re\big\{\big(\mathbf{h}_{k}^{(i)}-j\mathbf{x}_{\mathrm{c},k'}^{(i)}\!\!\ \herm\big)\nonumber \\
 & \quad\qquad\times\big(\mathbf{h}_{k}\herm+j\mathbf{x}_{\mathrm{c},k'}\big)\big\}+\|\mathbf{h}_{k}^{(i)}\!\!\ \herm+j\mathbf{x}_{\mathrm{c},k'}^{(i)}\|^{2}\big]\nonumber \\
 & \triangleq\bar{\mu}_{\mathrm{c},1,k,k'}\big(\mathbf{x}_{\mathrm{c},k'},\boldsymbol{\theta};\mathbf{x}_{\mathrm{c},k'}^{(i)},\boldsymbol{\theta}^{(i)}\big),\label{eq:muBar_c1_k_kPrime}\\
\bar{\wp}_{\mathrm{c},k,k'} & \geq\frac{1}{4}\big[\|\mathbf{h}_{k}\herm+j\mathbf{x}_{\mathrm{c},k'}\|^{2}-2\Re\big\{\big(\mathbf{h}_{k}^{(i)}+j\mathbf{x}_{\mathrm{c},k'}^{(i)}\!\!\ \herm\big)\nonumber \\
 & \quad\qquad\times\big(\mathbf{h}_{k}\herm-j\mathbf{x}_{\mathrm{c},k'}\big)\big\}+\|\mathbf{h}_{k}^{(i)}\!\!\ \herm-j\mathbf{x}_{\mathrm{c},k'}^{(i)}\|^{2}\big]\nonumber \\
 & \triangleq\bar{\mu}_{\mathrm{c},2,k,k'}\big(\mathbf{x}_{\mathrm{c},k'},\boldsymbol{\theta};\mathbf{x}_{\mathrm{c},k'}^{(i)},\boldsymbol{\theta}^{(i)}\big),\label{eq:muBar_c2_k_kPrime}\\
\wp_{\mathrm{t},k,m} & \geq\frac{1}{4}\big[\|\mathbf{h}_{k}\herm+\mathbf{x}_{\mathrm{t},m}\|^{2}-2\Re\big\{\big(\mathbf{h}_{k}^{(i)}-\mathbf{x}_{\mathrm{t},m}^{(i)}\!\!\ \herm\big)\nonumber \\
 & \quad\qquad\times\big(\mathbf{h}_{k}\herm-\mathbf{x}_{\mathrm{t},m}\big)\big\}+\|\mathbf{h}_{k}^{(i)}\!\!\ \herm-\mathbf{x}_{\mathrm{t},m}^{(i)}\|^{2}\big]\nonumber \\
 & \triangleq\mu_{\mathrm{r},1,k,m}\big(\mathbf{x}_{\mathrm{t},m},\boldsymbol{\theta};\mathbf{x}_{\mathrm{t},m}^{(i)},\boldsymbol{\theta}^{(i)}\big),\label{eq:mu_r1_k_m}\\
\wp_{\mathrm{t},k,m} & \geq\frac{1}{4}\big[\|\mathbf{h}_{k}\herm-\mathbf{x}_{\mathrm{t},m}\|^{2}-2\Re\big\{\big(\mathbf{h}_{k}^{(i)}+\mathbf{x}_{\mathrm{t},m}^{(i)}\!\!\ \herm\big)\nonumber \\
 & \quad\qquad\times\big(\mathbf{h}_{k}\herm+\mathbf{x}_{\mathrm{t},m}\big)\big\}+\|\mathbf{h}_{k}^{(i)}\!\!\ \herm+\mathbf{x}_{\mathrm{t},m}^{(i)}\|^{2}\big]\nonumber \\
 & \triangleq\mu_{\mathrm{r},2,k,m}\big(\mathbf{x}_{\mathrm{t},m},\boldsymbol{\theta};\mathbf{x}_{\mathrm{t},m}^{(i)},\boldsymbol{\theta}^{(i)}\big),\label{eq:mu_r2_k_m}\\
\bar{\wp}_{\mathrm{t},k,m} & \geq\frac{1}{4}\big[\|\mathbf{h}_{k}\herm-j\mathbf{x}_{\mathrm{t},m}\|^{2}-2\Re\big\{\big(\mathbf{h}_{k}^{(i)}-j\mathbf{x}_{\mathrm{t},m}^{(i)}\!\!\ \herm\big)\nonumber \\
 & \quad\qquad\times\big(\mathbf{h}_{k}\herm+j\mathbf{x}_{\mathrm{t},m}\big)\big\}+\|\mathbf{h}_{k}^{(i)}\!\!\ \herm+j\mathbf{x}_{\mathrm{t},m}^{(i)}\|^{2}\big]\nonumber \\
 & \triangleq\bar{\mu}_{\mathrm{r},1,k,m}\big(\mathbf{x}_{\mathrm{t},m},\boldsymbol{\theta};\mathbf{x}_{\mathrm{t},m}^{(i)},\boldsymbol{\theta}^{(i)}\big),\label{eq:muBar_r1_k_m}\\
\bar{\wp}_{\mathrm{t},k,m} & \geq\frac{1}{4}\big[\|\mathbf{h}_{k}\herm+j\mathbf{x}_{\mathrm{t},m}\|^{2}-2\Re\big\{\big(\mathbf{h}_{k}^{(i)}+j\mathbf{x}_{\mathrm{t},m}^{(i)}\!\!\ \herm\big)\nonumber \\
 & \quad\qquad\times\big(\mathbf{h}_{k}\herm-j\mathbf{x}_{\mathrm{t},m}\big)\big\}+\|\mathbf{h}_{k}^{(i)}\!\!\ \herm-j\mathbf{x}_{\mathrm{t},m}^{(i)}\|^{2}\big]\nonumber \\
 & \triangleq\bar{\mu}_{\mathrm{t},2,k,m}\big(\mathbf{x}_{\mathrm{t},m},\boldsymbol{\theta};\mathbf{x}_{\mathrm{t},m}^{(i)},\boldsymbol{\theta}^{(i)}\big).\label{eq:muBar_r2_k_m}
\end{align}
This completes the proof. 

\section{\protect\label{sec:proof-3}Proof of Theorem~\ref{thm-3}}

The constraint in~\eqref{eq:targetSINRC} for a given $k\in\mathcal{K}$ can be written as 
\begin{align}
 & \gamma_{\mathrm{t},k}\leq\Gamma_{\mathrm{t},k}\nonumber \\
\Rightarrow\  & \sigma_{\mathrm{t}}^{2}+\sum_{k'\in\mathcal{K}\setminus\{k\}}|\mathbf{g}_{\mathrm{t}}\mathbf{x}_{\mathrm{c},k'}|^{2}\nonumber \\
 & +\sum_{m\in\mathcal{M}}|\mathbf{g}_{\mathrm{t}}\mathbf{x}_{\mathrm{t},m}|^{2}+\sigma_{\mathrm{I}}^{2}\|\mathbf{g}_{\mathrm{R}}\boldsymbol{\Theta}\|^{2}\geq\frac{1}{\Gamma_{\mathrm{t},k}}|\mathbf{g}_{\mathrm{t}}\mathbf{x}_{\mathrm{c},k}|^{2}.\label{eq:SINRC-Tk-1}
\end{align}
Note that both the LHS and RHS of~\eqref{eq:SINRC-Tk-1} are neither convex nor concave, due to the coupling between $\mathbf{X}$ and $\boldsymbol{\theta}$. To proceed further, we need a concave lower bound on the LHS and a convex upper bound on the RHS of~\eqref{eq:SINRC-Tk-1}. Similar to~\eqref{eq:objLowerBound}, the former can be obtained with the help of~\eqref{eq:InEq1} and~\eqref{eq:InEq2}, resulting in the following: 
\begin{align}
 & \ \sigma_{\mathrm{t}}^{2}+\sum_{k'\in\mathcal{K}\setminus\{k\}}|\mathbf{g}_{\mathrm{t}}\mathbf{x}_{\mathrm{c},k'}|^{2}+\sum_{m\in\mathcal{M}}|\mathbf{g}_{\mathrm{t}}\mathbf{x}_{\mathrm{t},m}|^{2}+\sigma_{\mathrm{I}}^{2}\|\mathbf{g}_{\mathrm{R}}\boldsymbol{\Theta}\|^{2}\nonumber \\
\!\!\!\!\!\!\geq\  & \sigma_{\mathrm{t}}^{2}+\sum_{k'\in\mathcal{K}\setminus\{k\}}f_{k}(\mathbf{x}_{\mathrm{c},k'},\boldsymbol{\theta};\mathbf{x}_{\mathrm{c},k'}^{(i)},\boldsymbol{\theta}^{(i)})\nonumber \\
 & +\!\!\sum_{m\in\mathcal{M}}f_{m}(\mathbf{x}_{\mathrm{t},m},\boldsymbol{\theta};\mathbf{x}_{\mathrm{t},m}^{(i)},\boldsymbol{\theta}^{(i)})\!+\!\sigma_{\mathrm{I}}^{2}\!\sum_{n\in\mathscr{N}}\!\hat{f}_{n}(\theta_{n};\theta_{n}^{(i)}),\label{eq:SINRC-Tk-LHS}
\end{align}
which is jointly concave w.r.t. $\mathbf{X}$ and $\boldsymbol{\theta}$. Next, an upper bound on the RHS of~\eqref{eq:SINRC-Tk-1} can be analogous to~\eqref{eq:commSINR-1} . Hence, the constraint in~\eqref{eq:targetSINRC} can be transformed to
\begin{subequations}
\label{eq:secureInequalities}
\begin{align}
 & \sigma_{\mathrm{t}}^{2}+\sum_{k'\in\mathcal{K}\setminus\{K\}}f_{k}(\mathbf{x}_{\mathrm{c},k'},\boldsymbol{\theta};\mathbf{x}_{\mathrm{c},k'}^{(i)},\boldsymbol{\theta}^{(i)})\nonumber \\
 & +\!\!\sum_{m\in\mathcal{M}}f_{m}(\mathbf{x}_{\mathrm{t},m},\boldsymbol{\theta};\mathbf{x}_{\mathrm{t},m}^{(i)},\boldsymbol{\theta}^{(i)})\!+\!\sigma_{\mathrm{I}}^{2}\!\sum_{n\in\mathscr{N}}\!\hat{f}_{n}(\theta_{n};\theta_{n}^{(i)})\nonumber \\
 & \qquad\qquad\qquad\qquad\geq\frac{1}{\Gamma_{\mathrm{t},k}}\big(\tau_{\mathrm{c},k}^{2}+\bar{\tau}_{\mathrm{c},k}^{2}\big),\label{eq:secureSINRC}\\
 & \tau_{\mathrm{c},k}\geq\big|\Re\big\{\mathbf{g}_{\mathrm{t}}\mathbf{x}_{\mathrm{c},k}\big\}\big|,\label{eq:secureSINRC-real}\\
 & \bar{\tau}_{\mathrm{c},k}\geq\big|\Im\big\{\mathbf{g}_{\mathrm{t}}\mathbf{x}_{\mathrm{c},k}\big\}\big|.\label{eq:secureSINRC-imag}
\end{align}
\end{subequations}
It is noteworthy that if~\eqref{eq:targetSINRC} is feasible then so is~\eqref{eq:secureInequalities}, and vice versa. Following~\eqref{eq:tkl-1} \textendash{} \eqref{eq:muBar_r2_k_m}, we can transform the non-convex constraints in~\eqref{eq:secureSINRC-real} and~\eqref{eq:secureSINRC-imag} as follows:
\begin{subequations}
\label{eq:eta}
\begin{align}
\!\!\tau_{\mathrm{c},k} & \geq\frac{1}{4}\big[\|\mathbf{g}_{\mathrm{t}}\herm+\mathbf{x}_{\mathrm{c},k}\|^{2}-2\Re\big\{\big(\mathbf{g}_{\mathrm{t}}^{(i)}-\mathbf{x}_{\mathrm{c},k}^{(i)}\!\!\ \herm\big)\big(\mathbf{g}_{\mathrm{t}}\herm-\mathbf{x}_{\mathrm{c},k}\big)\big\}\nonumber \\
 & \quad+\|\mathbf{g}_{\mathrm{t}}^{(i)}\!\!\ \herm-\mathbf{x}_{\mathrm{c},k}^{(i)}\|^{2}\big]\triangleq\eta_{\mathrm{c},1,k}\big(\mathbf{x}_{\mathrm{c},k},\boldsymbol{\theta};\mathbf{x}_{\mathrm{c,}k}^{(i)},\boldsymbol{\theta}^{(i)}\big),\label{eq:eta_c1k}\\
\!\!\tau_{\mathrm{c},k} & \geq\frac{1}{4}\big[\|\mathbf{g}_{\mathrm{t}}\herm-\mathbf{x}_{\mathrm{c},k}\|^{2}-2\Re\big\{\big(\mathbf{g}_{\mathrm{t}}^{(i)}+\mathbf{x}_{\mathrm{c},k}^{(i)}\!\!\ \herm\big)\big(\mathbf{g}_{\mathrm{t}}\herm+\mathbf{x}_{\mathrm{c},k}\big)\big\}\nonumber \\
 & \quad+\|\mathbf{g}_{\mathrm{t}}^{(i)}\!\!\ \herm+\mathbf{x}_{\mathrm{c},k}^{(i)}\|^{2}\big]\triangleq\eta_{\mathrm{c,2,}k}\big(\mathbf{x}_{\mathrm{c},k},\boldsymbol{\theta};\mathbf{x}_{\mathrm{c},k}^{(i)},\boldsymbol{\theta}^{(i)}\big),\label{eq:eta_c2k}\\
\!\!\bar{\tau}_{\mathrm{c},k} & \geq\frac{1}{4}\big[\|\mathbf{g}_{\mathrm{t}}\herm-j\mathbf{x}_{\mathrm{c},k}\|^{2}-2\Re\big\{\big(\mathbf{g}_{\mathrm{t}}^{(i)}-j\mathbf{x}_{\mathrm{c},k}^{(i)}\!\!\ \herm\big)\big(\mathbf{g}_{\mathrm{t}}\herm+j\mathbf{x}_{\mathrm{c},k}\big)\big\}\nonumber \\
 & \quad+\|\mathbf{g}_{\mathrm{t}}^{(i)}\!\!\ \herm\!+\!j\mathbf{x}_{\mathrm{c},k}^{(i)}\|^{2}\big]\triangleq\bar{\eta}_{\mathrm{c},1,k}\big(\mathbf{x}_{\mathrm{c},k},\boldsymbol{\theta};\mathbf{x}_{\mathrm{c},k}^{(i)},\boldsymbol{\theta}^{(i)}\big),\label{eq:etaBar_c1k}\\
\!\!\bar{\tau}_{\mathrm{c},k} & \geq\frac{1}{4}\big[\|\mathbf{g}_{\mathrm{t}}\herm+j\mathbf{x}_{\mathrm{c},k}\|^{2}-2\Re\big\{\big(\mathbf{g}_{\mathrm{t}}^{(i)}+j\mathbf{x}_{\mathrm{c},k}^{(i)}\!\!\ \herm\big)\big(\mathbf{g}_{\mathrm{t}}\herm-j\mathbf{x}_{\mathrm{c},k}\big)\big\}\nonumber \\
 & \quad+\|\mathbf{g}_{\mathrm{t}}^{(i)}\!\!\ \herm\!-\!j\mathbf{x}_{\mathrm{c},k}^{(i)}\|^{2}\big]\!\triangleq\bar{\eta}_{\mathrm{c},2,k}\big(\mathbf{x}_{\mathrm{c},k},\boldsymbol{\theta};\mathbf{x}_{\mathrm{c},k}^{(i)},\boldsymbol{\theta}^{(i)}\big).\label{eq:etaBar_c2k}
\end{align}
\end{subequations}
This concludes the proof. 

\section{\protect\label{sec:proof-4}Proof of Theorem~\ref{thm-4}}

From~\eqref{eq:powerConsumption}, it can be noted that the LHS of~\eqref{eq:TPC} is non-convex, and therefore an equivalent reformulation of~\eqref{eq:TPC} can be given by 
\begin{subequations}
\label{eq:TPC-1}
\begin{align}
P_{\mathrm{\max}}\geq & \ \sum_{k\in\mathcal{K}}\|\mathbf{x}_{\mathrm{c},k}\|^{2}+\sum_{m\in\mathcal{M}}\|\mathbf{x}_{\mathrm{t},m}\|^{2}\nonumber \\
 & +\sum_{k\in\mathcal{K}}\sum_{n\in\mathscr{N}}(\varkappa_{\mathrm{c},k,n}^{2}+\bar{\varkappa}_{\mathrm{c},k,n}^{2})\nonumber \\
 & +\sum_{m\in\mathcal{M}}\sum_{n\in\mathscr{N}}(\varkappa_{\mathrm{t},m,n}^{2}+\bar{\varkappa}_{\mathrm{t},m,n}^{2})+\sigma_{\mathrm{I}}^{2}\sum_{n\in\mathscr{N}}|\theta_{n}|^{2},\label{eq:TPC-1-1}\\
\varkappa_{\mathrm{c},k,n}\geq & \ |\Re\{\theta_{n}\mathbf{g}_{n}\mathbf{x}_{\mathrm{c},k}\}|\ \forall k\in\mathcal{K},n\in\mathscr{N},\label{eq:TPC-1-2}\\
\bar{\varkappa}_{\mathrm{c},k,n}\geq & \ |\Im\{\theta_{n}\mathbf{g}_{n}\mathbf{x}_{\mathrm{c},k}\}|\ \forall k\in\mathcal{K},n\in\mathscr{N},\label{eq:TPC-1-3}\\
\varkappa_{\mathrm{t},m,n}\geq & \ |\Re\{\theta_{n}\mathbf{g}_{n}\mathbf{x}_{\mathrm{t},m}\}|\ \forall m\in\mathcal{M},n\in\mathscr{N},\label{eq:TPC-1-4}\\
\bar{\varkappa}_{\mathrm{t},m,n}\geq & \ |\Im\{\theta_{n}\mathbf{g}_{n}\mathbf{x}_{\mathrm{t},m}\}|\ \forall m\in\mathcal{M},n\in\mathscr{N},\label{eq:TPC-1-5}
\end{align}
\end{subequations}
where $\mathbf{g}_{n}\in\mathbb{C}^{1\times L}$ represents the $n$-th row of $\mathbf{G}$. It is straightforward to see that if~\eqref{eq:TPC} is feasible, then so is~\eqref{eq:TPC-1} and vice versa; this means that the optimality of~\eqref{eq:TPC} remains unaffected after the reformulation. Furthermore, the constraints in~\eqref{eq:TPC-1-2} \textendash{} \eqref{eq:TPC-1-5} can be further represented as 
\begin{subequations}
\label{eq:varKappa}
\begin{align}
\varkappa_{\mathrm{c},k,n} & \geq\frac{1}{4}\big[\|\mathbf{g}_{n}\herm\theta_{n}\herm+\mathbf{x}_{\mathrm{c},k}\|^{2}-2\Re\big\{\big(\theta_{n}^{(i)}\mathbf{g}_{n}-\mathbf{x}_{\mathrm{c},k}^{(i)}\ \!\herm\big)\nonumber \\
 & \ \times\big(\mathbf{g}_{n}\herm\theta_{n}\herm-\mathbf{x}_{\mathrm{c},k}\big)\big\}+\|\mathbf{g}_{n}\herm\theta_{n}^{(i)}\ \!\herm-\mathbf{x}_{\mathrm{c},k}^{(i)}\|^{2}\big]\nonumber \\
 & \triangleq\varpi_{\mathrm{c},1,k,n}(\mathbf{x}_{\mathrm{c},k},\theta_{n};\mathbf{x}_{\mathrm{c},k}^{(i)},\theta_{n}^{(i)}),\label{eq:eta_c1kn}\\
\varkappa_{\mathrm{c},k,n} & \geq\frac{1}{4}\big[\|\mathbf{g}_{n}\herm\theta_{n}\herm-\mathbf{x}_{\mathrm{c},k}\|^{2}-2\Re\big\{\big(\theta_{n}^{(i)}\mathbf{g}_{n}+\mathbf{x}_{\mathrm{c},k}^{(i)}\ \!\herm\big)\nonumber \\
 & \ \times\big(\mathbf{g}_{n}\herm\theta_{n}\herm+\mathbf{x}_{\mathrm{c},k}\big)\big\}+\|\mathbf{g}_{n}\herm\theta_{n}^{(i)}\ \!\herm+\mathbf{x}_{\mathrm{c},k}^{(i)}\|^{2}\big]\nonumber \\
 & \triangleq\varpi_{\mathrm{c},2,k,n}(\mathbf{x}_{\mathrm{c},k},\theta_{n};\mathbf{x}_{\mathrm{c},k}^{(i)},\theta_{n}^{(i)}),\label{eq:eta_c2kn}\\
\bar{\varkappa}_{\mathrm{c},k,n} & \geq\frac{1}{4}\big[\|\mathbf{g}_{n}\herm\theta_{n}\herm-j\mathbf{x}_{\mathrm{c},k}\|^{2}-2\Re\big\{\big(\theta_{n}^{(i)}\mathbf{g}_{n}-j\mathbf{x}_{\mathrm{c},k}^{(i)}\ \!\herm\big)\nonumber \\
 & \ \times\big(\mathbf{g}_{n}\herm\theta_{n}\herm+j\mathbf{x}_{\mathrm{c},k}\big)\big\}+\|\mathbf{g}_{n}\herm\theta_{n}^{(i)}\ \!\herm+j\mathbf{x}_{\mathrm{c},k}^{(i)}\|^{2}\big]\nonumber \\
 & \triangleq\bar{\varpi}_{\mathrm{c},1,k,n}(\mathbf{x}_{\mathrm{c},k},\theta_{n};\mathbf{x}_{\mathrm{c},k}^{(i)},\theta_{n}^{(i)}),\label{eq:etaBar_c1kn}\\
\bar{\varkappa}_{\mathrm{c},k,n} & \geq\frac{1}{4}\big[\|\mathbf{g}_{n}\herm\theta_{n}\herm+j\mathbf{x}_{\mathrm{c},k}\|^{2}-2\Re\big\{\big(\theta_{n}^{(i)}\mathbf{g}_{n}+j\mathbf{x}_{\mathrm{c},k}^{(i)}\ \!\herm\big)\nonumber \\
 & \ \times\big(\mathbf{g}_{n}\herm\theta_{n}\herm-j\mathbf{x}_{\mathrm{c},k}\big)\big\}+\|\mathbf{g}_{n}\herm\theta_{n}^{(i)}\ \!\herm-j\mathbf{x}_{\mathrm{c},k}^{(i)}\|^{2}\big]\nonumber \\
 & \triangleq\bar{\varpi}_{\mathrm{c},2,k,n}(\mathbf{x}_{\mathrm{c},k},\theta_{n};\mathbf{x}_{\mathrm{c},k}^{(i)},\theta_{n}^{(i)}),\label{eq:etaBar_c2kn}\\
\varkappa_{\mathrm{t},m,n} & \geq\frac{1}{4}\big[\|\mathbf{g}_{n}\herm\theta_{n}\herm+\mathbf{x}_{\mathrm{t},m}\|^{2}-2\Re\big\{\big(\theta_{n}^{(i)}\mathbf{g}_{n}-\mathbf{x}_{\mathrm{t},m}^{(i)}\ \!\herm\big)\nonumber \\
 & \ \times\big(\mathbf{g}_{n}\herm\theta_{n}\herm-\mathbf{x}_{\mathrm{t},m}\big)\big\}+\|\mathbf{g}_{n}\herm\theta_{n}^{(i)}\ \!\herm-\mathbf{x}_{\mathrm{t},m}^{(i)}\|^{2}\big]\nonumber \\
 & \triangleq\varpi_{\mathrm{t},1,m,n}(\mathbf{x}_{\mathrm{t},m},\theta_{n};\mathbf{x}_{\mathrm{t},m}^{(i)},\theta_{n}^{(i)}),\label{eq:eta_t1mn}\\
\varkappa_{\mathrm{t},m,n} & \geq\frac{1}{4}\big[\|\mathbf{g}_{n}\herm\theta_{n}\herm-\mathbf{x}_{\mathrm{t},m}\|^{2}-2\Re\big\{\big(\theta_{n}^{(i)}\mathbf{g}_{n}+\mathbf{x}_{\mathrm{t},m}^{(i)}\ \!\herm\big)\nonumber \\
 & \ \times\big(\mathbf{g}_{n}\herm\theta_{n}\herm+\mathbf{x}_{\mathrm{t},m}\big)\big\}+\|\mathbf{g}_{n}\herm\theta_{n}^{(i)}\ \!\herm+\mathbf{x}_{\mathrm{t},m}^{(i)}\|^{2}\big]\nonumber \\
 & \triangleq\varpi_{\mathrm{t},2,m,n}(\mathbf{x}_{\mathrm{t},m},\theta_{n};\mathbf{x}_{\mathrm{t},m}^{(i)},\theta_{n}^{(i)}),\label{eq:eta_t2mn}\\
\bar{\varkappa}_{\mathrm{t},m,n} & \geq\frac{1}{4}\big[\|\mathbf{g}_{n}\herm\theta_{n}\herm-j\mathbf{x}_{\mathrm{t},m}\|^{2}-2\Re\big\{\big(\theta_{n}^{(i)}\mathbf{g}_{n}-j\mathbf{x}_{\mathrm{t},m}^{(i)}\ \!\herm\big)\nonumber \\
 & \ \times\big(\mathbf{g}_{n}\herm\theta_{n}\herm+j\mathbf{x}_{\mathrm{t},m}\big)\big\}+\|\mathbf{g}_{n}\herm\theta_{n}^{(i)}\ \!\herm+j\mathbf{x}_{\mathrm{t},m}^{(i)}\|^{2}\big]\nonumber \\
 & \triangleq\bar{\varpi}_{\mathrm{t},1,m,n}(\mathbf{x}_{\mathrm{t},m},\theta_{n};\mathbf{x}_{\mathrm{t},m}^{(i)},\theta_{n}^{(i)}),\label{eq:etaBar_t1mn}\\
\bar{\varkappa}_{\mathrm{t},m,n} & \geq\frac{1}{4}\big[\|\mathbf{g}_{n}\herm\theta_{n}\herm+j\mathbf{x}_{\mathrm{t},m}\|^{2}-2\Re\big\{\big(\theta_{n}^{(i)}\mathbf{g}_{n}+j\mathbf{x}_{\mathrm{t},m}^{(i)}\ \!\herm\big)\nonumber \\
 & \ \times\big(\mathbf{g}_{n}\herm\theta_{n}\herm-j\mathbf{x}_{\mathrm{t},m}\big)\big\}+\|\mathbf{g}_{n}\herm\theta_{n}^{(i)}\ \!\herm-j\mathbf{x}_{\mathrm{t},m}^{(i)}\|^{2}\big]\nonumber \\
 & \triangleq\bar{\varpi}_{\mathrm{t},2,m,n}(\mathbf{x}_{\mathrm{t},m},\theta_{n};\mathbf{x}_{\mathrm{t},m}^{(i)},\theta_{n}^{(i)}).\label{eq:etaBar_t2mn}
\end{align}
\end{subequations}
The proof is now complete. \ifCLASSOPTIONcaptionsoff
\newpage
\fi
\bibliographystyle{IEEEtran}
\bibliography{Paper-TW-Sep-24-2090.R2}


\begin{IEEEbiography}[{\includegraphics[width=1in,height=1.25in,clip,keepaspectratio]{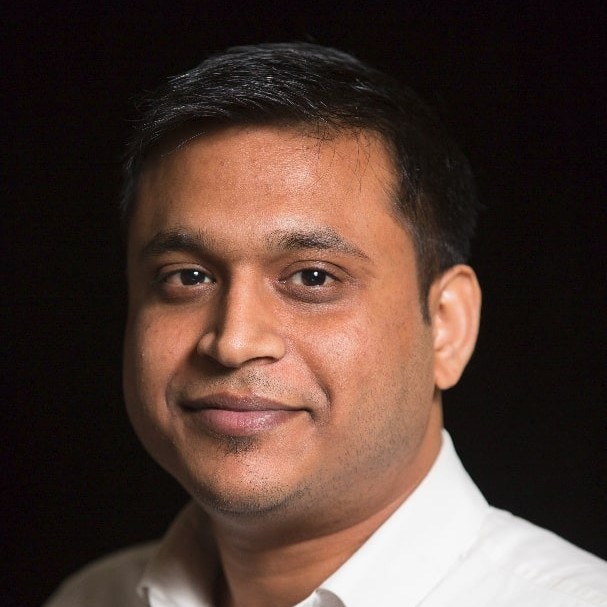}}]{Vaibhav Kumar} (Member, IEEE) received his B.E. degree in Electronics and Telecommunications Engineering from CSVTU, Bhilai, India, in 2012. He completed his M.Tech. in Electronics and Communication Engineering from The LNM Institute of Information Technology, Jaipur, India, in 2015, and earned his Ph.D. in Electronic Engineering from University College Dublin (UCD), Ireland, in 2020. From 2012 to 2013, he served as a Lecturer in the Department of Biomedical Engineering at the National Institute of Technology, Raipur, India. Between 2015 and 2016, he worked as a Project Associate on a project titled ``Mobile Broadband Service Support over Cognitive Radio Networks,'' funded by Media Lab Asia, Government of India. In early 2019, he was a Visiting Research Student at the Indian Institute of Technology Delhi under the Erasmus+ ICM research program.
	
	From 2020 to 2023, he was a Postdoctoral Research Fellow in the School of Electrical and Electronic Engineering at UCD. He also held an Adjunct Lecturer position at Beijing-Dublin International College from September to December 2020. Since September 2023, he has been working as a Postdoctoral Associate at New York University Abu Dhabi (NYUAD), United Arab Emirates. Dr. Kumar received the Best Paper Award at the Communication \& Information System Security Symposium of IEEE GLOBECOM 2021 in Madrid, Spain, and the Exemplary Reviewer Award from \textsc{IEEE Communications Letters} in 2022. He is currently an Area Editor for the Physical Communication (Elsevier). His research interests lie in the areas of wireless communications and signal processing.
\end{IEEEbiography}

\begin{IEEEbiography}[{\includegraphics[width=1in,height=1.25in,clip,keepaspectratio]{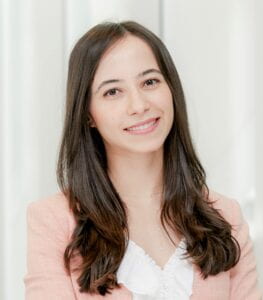}}]{Marwa Chafii} (Senior Member, IEEE) received her Ph.D. degree in Electrical Engineering in 2016 and her Master's degree in Advanced Wireless Communication Systems (SAR) in 2013, both from CentraleSup\'elec, France. Between 2014 and 2016, she was a visiting researcher at Poznan University of Technology (Poland), the University of York (UK), Yokohama National University (Japan), and the University of Oxford (UK). She joined the Technical University of Dresden, Germany, in 2018 as a research group leader, and ENSEA, France, in 2019 as an Associate Professor, where she held a Chair of Excellence on Artificial Intelligence from the CY Initiative. Since September 2021, she has been an Associate Professor at New York University (NYU) Abu Dhabi and NYU WIRELESS at NYU Tandon School of Engineering. Her research interests include advanced waveform design, machine learning for wireless communications, and indoor localization.
	
	She received the IEEE ComSoc Best Young Researcher Award for the Europe, Middle East, and Africa (EMEA) region, the IEEE ComSoc Best Young Professional Practitioner Award, the IEEE Glavieux Award, and the Best Ph.D. Thesis Prize in France in the fields of Signal, Image, and Vision. In 2020, she was selected among the Top 10 Rising Stars in Computer Networking and Communications by N2Women. She served as an Associate Editor for \textsc{IEEE Communications Letters} from 2019 to 2021, where she received the Best Editor Award in 2020. From 2018 to 2021, she was the Research Lead at Women in AI. She is currently an Associate Editor for \textsc{IEEE Transactions on Communications}, serves as Vice-Chair of the IEEE ComSoc Emerging Technologies Initiative (ETI) on Machine Learning for Communications, and leads the Education Working Group of the IEEE ComSoc ETI on Integrated Sensing and Communications.
\end{IEEEbiography}

\end{document}